\documentclass[11pt,a4paper]{article}

\usepackage{amssymb}
\usepackage{amsthm}
\usepackage{amsmath}            %includes a package
\usepackage{latexsym}
\usepackage{epsfig}
\usepackage{amscd}
\usepackage{graphicx}
\usepackage[dvips]{color}
\renewcommand{\proof}[1]{\noindent \normalfont{\textbf{Proof.} \  \  #1 \hfill $\Box$} \par}

\newcommand{\ben}{\begin{equation}}     %equation
\newcommand{\eeqn}{\end{equation}}
\newcommand{\bey}{\begin{eqnarray}}
\newcommand{\eey}{\end{eqnarray}}

\newtheorem{thm}{Theorem}[section]

\newtheorem{lemma}[thm]{Lemma}

\begin{document}

\bibliographystyle{plain}

\pagenumbering{arabic}
\setcounter{page}{1}

\title{Hebbian inspecificity in the Oja model}
\author{}
\maketitle

\begin{abstract}
\small{Recent work on Long Term Potentiation in brain slices shows that Hebb's rule is not completely synapse-specific, probably due to intersynapse diffusion of calcium or other factors. We extend the classical Oja unsupervised model of learning by a single linear neuron to include Hebbian inspecificity, by introducing an error matrix ${\bf E}$, which expresses possible crosstalk between updating at different connections. We show the modified algorithm converges to the leading eigenvector of the matrix ${\bf EC}$, where ${\bf C}$ is the input covariance matrix. When there is no inspecificity (i.e. ${\bf E}$ is the identity matrix), this gives the classical result of convergence to the first principal component of the input distribution (PC1). We then study the outcome of learning using different versions of ${\bf E}$. In the most biologically plausible case, ``error-onto-all'', arising when there are no intrinsically privileged connections, ${\bf E}$ has diagonal elements $Q$ and off-diagonal elements $(1-Q)/(n-1)$, where $Q$, the quality, is expected to decrease with the number of inputs $n$. With reasonable assumptions about the biophysics of Long Term Potentiation we can take $Q=(1-b)^n$ (in a discrete model) or $Q=1/(bn+1)$ (in a continuous model), where $b$ is a single synapse inaccuracy parameter that reflects synapse density, calcium diffusion, etc. We analyze this error-onto-all case in detail, for both uncorrelated and correlated inputs. We study the dependence of the angle $\theta$ between PC1 and the leading eigenvector of ${\bf EC}$ on $b$, $n$ and the amount of input activity or correlation. (We do this analytically and using Matlab calculations.) We find that $\theta$ increases (learning becomes gradually less useful) with increases in $b$, particularly for intermediate (i.e. biologically-realistic) correlation strength, although some useful learning always occurs up to the trivial limit $Q=1/n$. We discuss the relation of our results to Hebbian unsupervised learning in the brain.}
\end{abstract}

\vspace{10mm}

\section{Introduction}
\indent

Various brain structures such as the neocortex are believed to use
unsupervised synaptic learning to form neural representations that
capture and exploit statistical regularities of an animal's
world. Most neural models of unsupervised learning use some form of
Hebb rule to update synaptic connections. Typically, this rule is
implemented by updating a connection according to the product of the
input and output firing rates. Other forms of the update rule are
sometimes used, but they are still typically local and activity
dependent, and often Hebbian in the sense that they depend on both
input and output activity. Biological networks may also use
spike-timing dependent rules, but these are also Hebbian in the
sense that they depend on the relative timing of pre- and
postsynaptic spiking. The key element in Hebbian learning is that
the update should depend on the extent to which the input appears to
``take part in'' firing the output (~\cite{Hebb}; we added ``appears''
to emphasize that a single neural connection knows nothing about
actual causation, and merely responds to a statistical coincidence
of pre- and postsynaptic spikes).

We are interested in the possibility that if the Hebb rule is not
completely local (in the sense there might be some, possibly very
weak, dependence of the local update on activity at other
connections) unsupervised learning might fail catastrophically, not
only preventing new learning, but wiping out previous learning. We
have proposed that the basic task of the neocortex is to avoid such
hypothetical learning catastrophes~\cite{Cox}~\cite{Adams1}. In this paper we modify a classical model of unsupervised learning, the Oja single neuron
Principal Component Analyzer~\cite{Oja1}, to include Hebbian inaccuracy. By ``inspecificity", or ``inaccuracy'', we mean that part of the local update calculated using a Hebb rule (for example, proportional to the product of input and output firing rates) is assigned to connections other than the one at which the product was calculated. We also refer to this
postulated  nonlocality as ``leakage'', ``crosstalk'', or simply ``error''. Some other papers on this topic have appeared~\cite{Botelho}, and
in the Discussion we will try to clarify the relationship between
these various studies. We will conclude that while the modified Oja
model, and perhaps others that are only sensitive to second-order
statistics, do not show a true error catastrophe at finite network
size, their behaviour gives important clues to understanding the
difficulties that brains might encounter in learning higher-order
statistics.

Recent experimental work has shown that long term potentiation
(LTP), a biological manifestation of the Hebb rule, is indeed not
completely synapse specific~\cite{Engert1}~\cite{Schuman}~\cite{Bi}~\cite{Tao}. For example, Engert and
Bonhoeffer have shown that LTP induced at a local set of connections
on a CA1 pyramidal cell ``spills over'' to induce LTP at a nearby
set of inactive connections. In earlier work using less refined
methods, it had been concluded that LTP was synapse-specific~\cite{Andersen}~\cite{Levy}. Even in the Engert -Bonhoeffer experiments~\cite{Engert1}, it is likely that, because the ``pairing'' method used to induce LTP was rather
crude, the inspecificity was far greater than would ever be actually
seen in an awake brain. More recent work has shown that at least one
type of Hebbian inspecificity, induced by theta burst stimulation of
retinotectal connections, reflects dendritic spread of calcium~\cite{Tao}. Even more recent LTP experiments at single synapses have shown that, while LTP is only expressed locally~\cite{Matsuzaki}, the threshold for LTP induced at neighboring synapses is reduced~\cite{Harvey}. Thus, some degree of Hebbian inspecificity is probably inevitable, and its effects on learning need to be evaluated.

\section{Overview}
\noindent

   Here we briefly review the classical Oja model~\cite{Oja1}~\cite{Oja2}~\cite{Diamantaras}~\cite{Herz} and define terms as background to the new analysis.
   The model network consists of a single output neuron receiving $n$ signals $x_{1},x_{2},...,x_{n}$ from a set of $n$ input neurons via connections of corresponding strengths $\omega_{1},...,\omega_{n}$ (see Figure 1).

The resulting output $y$ is defined as the weighted sum of the inputs:
 $$y=\sum_{i=1}^{n}x_{i}\omega_{i}$$

\noindent The input column vector ${\bf x}=(x_{1}...x_{n})^{T}$ is randomly drawn from a probability distribution ${\cal{P}}({\bf x}), \; {\bf x} \in \mathbb{R}^{n}$ (where $^{T}$ denotes transposition of vectors).

   In accordance to Hebb's postulate of learning, a synaptic weight $\omega_{i}$ will strengthen proportionally with the product of $x_{i}$ and $y$:
   $$\omega_{i}(t+1) = \omega_{i}(t) + \gamma y(t) x_{i}(t)$$

Here $\gamma$ is a time independent learning rate and the argument
$t$ represents the dependence on time (or on the input draw). The relation between this formulation and neural processes such as LTP is considered in the Discussion.

   Oja~\cite{Oja1} modified this by normalizing  the weight vector $\mbox{\boldmath $\omega$}$ with respect to the Euclidean metric on $\mathbb{R}^{n}$:
$$\omega_{i}(t+1) = \frac{\omega_{i}(t) + \gamma y(t) x_{i}(t)}{\parallel \mbox{\boldmath $\omega$}(t) + \gamma y(t) {\bf x}(t) \parallel}$$
Expanding in Taylor series with respect to $\gamma$ and ignoring the
${\cal O}(\gamma^{2})$ term for $\gamma$ sufficiently small, the
result is:

$$\mbox{\boldmath $\omega$}(t+1) = \mbox{\boldmath $\omega$}(t)+\gamma y(t)[{\bf x}(t)- y(t)\mbox{\boldmath $\omega$}(t)]$$

\noindent Henceforth, we omit the variable $t$ whenever there is no ambiguity. The equation can be then rewritten as:

$$\mbox{\boldmath $\omega$}(t+1) = \mbox{\boldmath $\omega$} + \gamma \left[ {\bf xx}^{T}\mbox{\boldmath $\omega$}- \left( \mbox{\boldmath $\omega$}^{T}{\bf xx}^{T}\mbox{\boldmath $\omega$} \right) \mbox{\boldmath $\omega$} \right]$$

   Consider the covariance matrix of the distribution ${\cal{P}}({\bf x})$, defined by $C=\langle {\bf xx}^T \rangle=\langle {\bf x}(t){\bf x}^T(t) \rangle$. Clearly ${\bf C}$ is symmetric and semipositive definite. With the following additional assumptions:

   $\bullet$ the learning process is slow enough for $\mbox{\boldmath $\omega$}$ to be treated as stationary;

   $\bullet$ ${\bf x}(t)$ and $\mbox{\boldmath $\omega$}(t)$ are statistically independent

\noindent we can take conditional expectation over ${\cal{P}}({\bf x})$
and rewrite the learning rule as:

$$\displaystyle{\langle \mbox{\boldmath $\omega$}(t+1) \vert \mbox{\boldmath $\omega$}(t) \rangle = {\bf w} + \gamma \left[ {\bf C w} - \left( {\bf w}^{T}{\bf C w} \right) \right]}$$

Oja concluded that, if $\mbox{\boldmath $\omega$}(t)$ converges as $t \rightarrow \infty$, the limit is expected to be one of the two opposite normalized eigenvectors corresponding to the maximal eigenvalue of ${\bf C}$ (i.e., the ``principal component" of the matrix ${\bf C}$~\cite{Oja1}~\cite{Oja2}).\\

The general form of the rule, which we use throughout this paper, allows elements of ${\bf x}$ and \mbox{\boldmath $\omega$} to be negative. However, there is a biologically interesting special case: all components of ${\bf x}$ (corresponding to firing rates) and \mbox{\boldmath $\omega$} (corresponding to synaptic strengths) are always positive, so the first term corresponds to LTP, and the second term corresponds to long term depression (LTD) (see Discussion).

The operation of the Oja rule can be understood intuitively in the following way. Each new input vector twists the current weight vector, which is constrained to lie close to the unit hypersphere surface, towards itself. If all the twists had magnitudes proportional to the corresponding input vector, the final weight vector would lie in the direction of the mean of the input distribution. However, in the Oja rule the magnitude of the twist also depends on the output: the influence of input vectors that are closer in direction to the current weight vector is magnified, since their dot product with that weight vector is larger; thus as the weight vector's direction gets closer to the first principal component direction, those arriving input patterns that are closer to PC1 produce larger twists than those that are further away. Thus the different PCs do not all contribute to the final outcome; instead, only the largest PC wins. We will see that the effect of error is to moderate this ``winner-take-all'' behavior; inaccurate learning leads to incomplete victory, such that the learned weight vector is a linear combination of the eigenvectors of ${\bf C}$. However, as long as learning shows some degree of specificity, the learned weight vector will always be closer to the first PC than to any others; in this sense learning continues to be useful, at least  for Gaussian inputs, since it allows a more useful scalar representation than merely randomly selecting one of the input pattern elements.\\

   To introduce inspecificity into the learning equation, we assume that, on average, only a fraction $Q$ of the intended update reaches the appropriate connection, the remaining fraction $1-Q$ being distributed amongst the other connections according to a defined and biologically plausible rule. The actual update at a given connection thus includes contributions from erroneous or innacurate updates from other connections. The erroneous updating process is formally described by a possibly time-dependent error matrix  $\mbox{\boldmath ${\cal E}$}=\mbox{\boldmath ${\cal E}$}(t)$, independent of the inputs, whose elements, which depend, on average, on $Q$, reflect at each time step $t$ the fractional contribution that the activity across weight $\omega_{i}$ makes to the update of $\omega_{j}$. Then Hebb's rule changes into the normalized version of

$$\omega_{i}(t+1) = \omega_{i} + \gamma y[\mbox{\boldmath ${\cal E}$}{\bf x}]_{i}$$

\noindent which, after normalization and linearization with respect to $\gamma$, becomes:

$$\omega_{i}(t+1) = \omega_{i} + \gamma y([\mbox{\boldmath ${\cal E}$}{\bf x}]_{i}- y \omega_{i})$$

\noindent Taking conditional expectation of both sides and rewriting the equation in matrix form leads to:

$$\langle \mbox{\boldmath $\omega$}(t+1) \vert \mbox{\boldmath $\omega$}(t) \rangle = {\bf w} + \gamma \left[ {\bf ECw}- \left( {\bf w}^{T}{\bf Cw} \right) {\bf w} \right]$$

\noindent where we defined ${\bf w}=\langle \mbox{\boldmath $\omega$} \rangle$ and ${\bf E}= \langle \mbox{\boldmath ${\cal E}$} \rangle$. ${\bf E}$ is then a symmetric circulant matrix; in the zero error case of $Q=1$, ${\bf E}$ would become the identity matrix.

\begin{figure}[h!]
\begin{center}
\includegraphics[scale=.55]{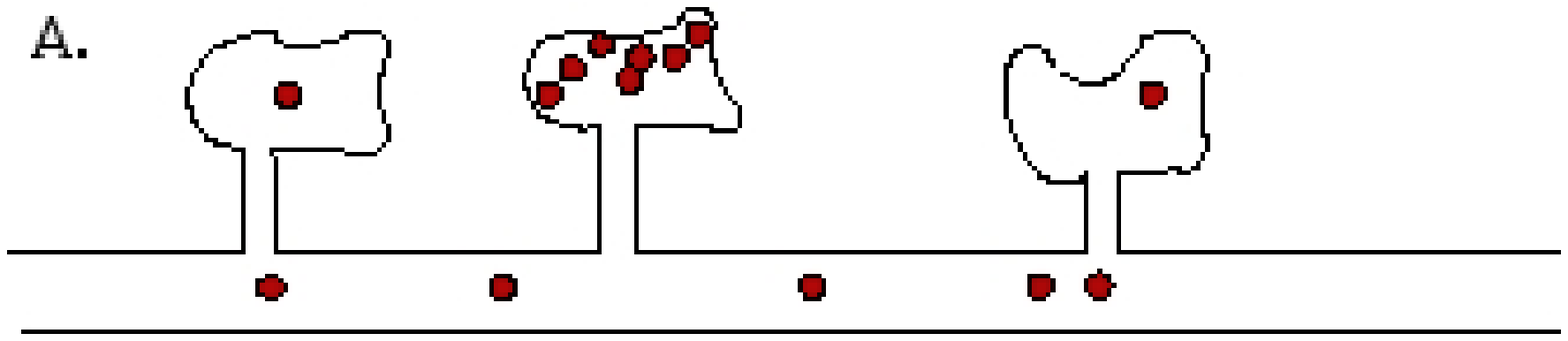}
\end{center}
\vspace{2mm}
\begin{center}
\includegraphics[scale=0.3]{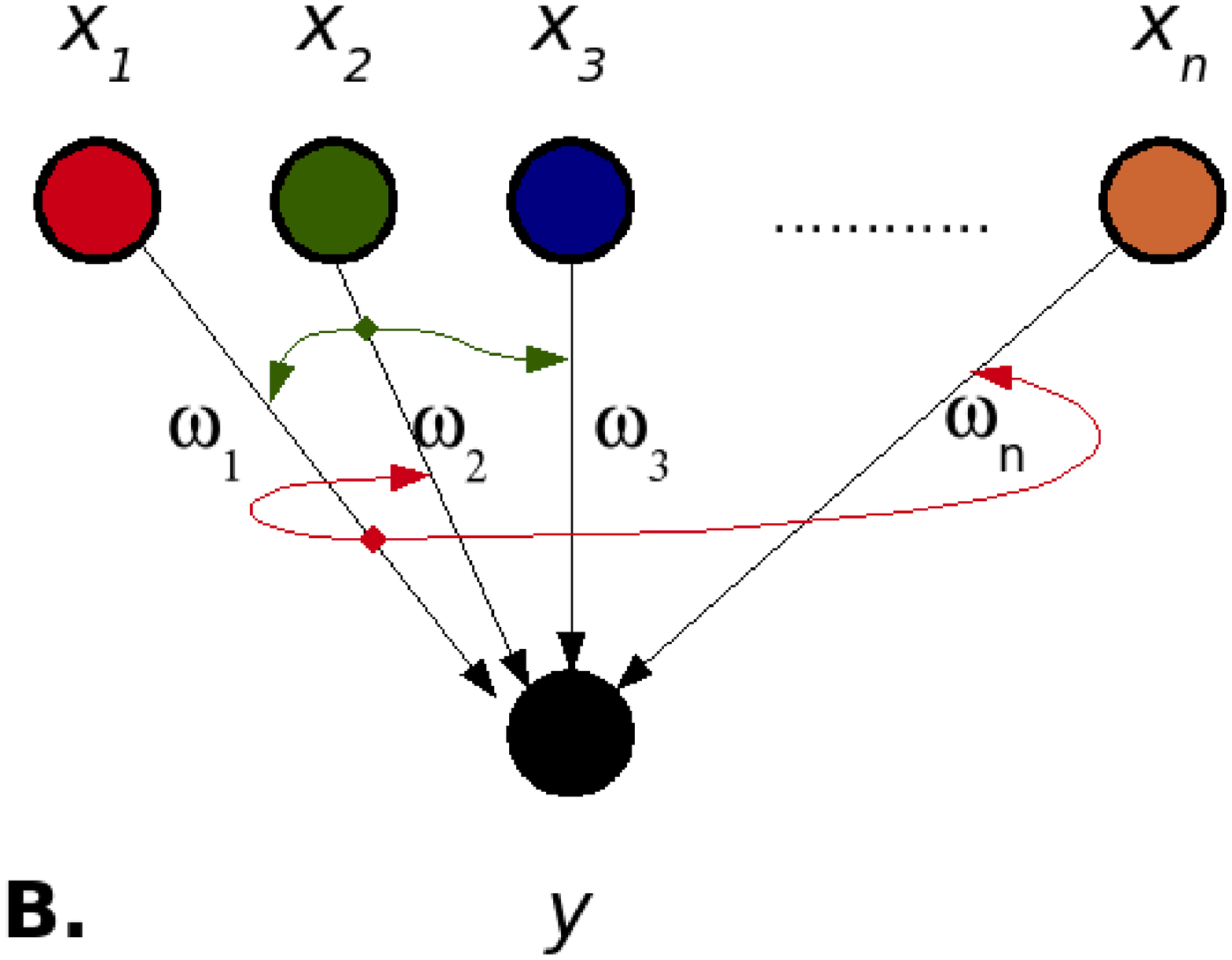} \quad \quad \quad
\includegraphics[scale=0.3]{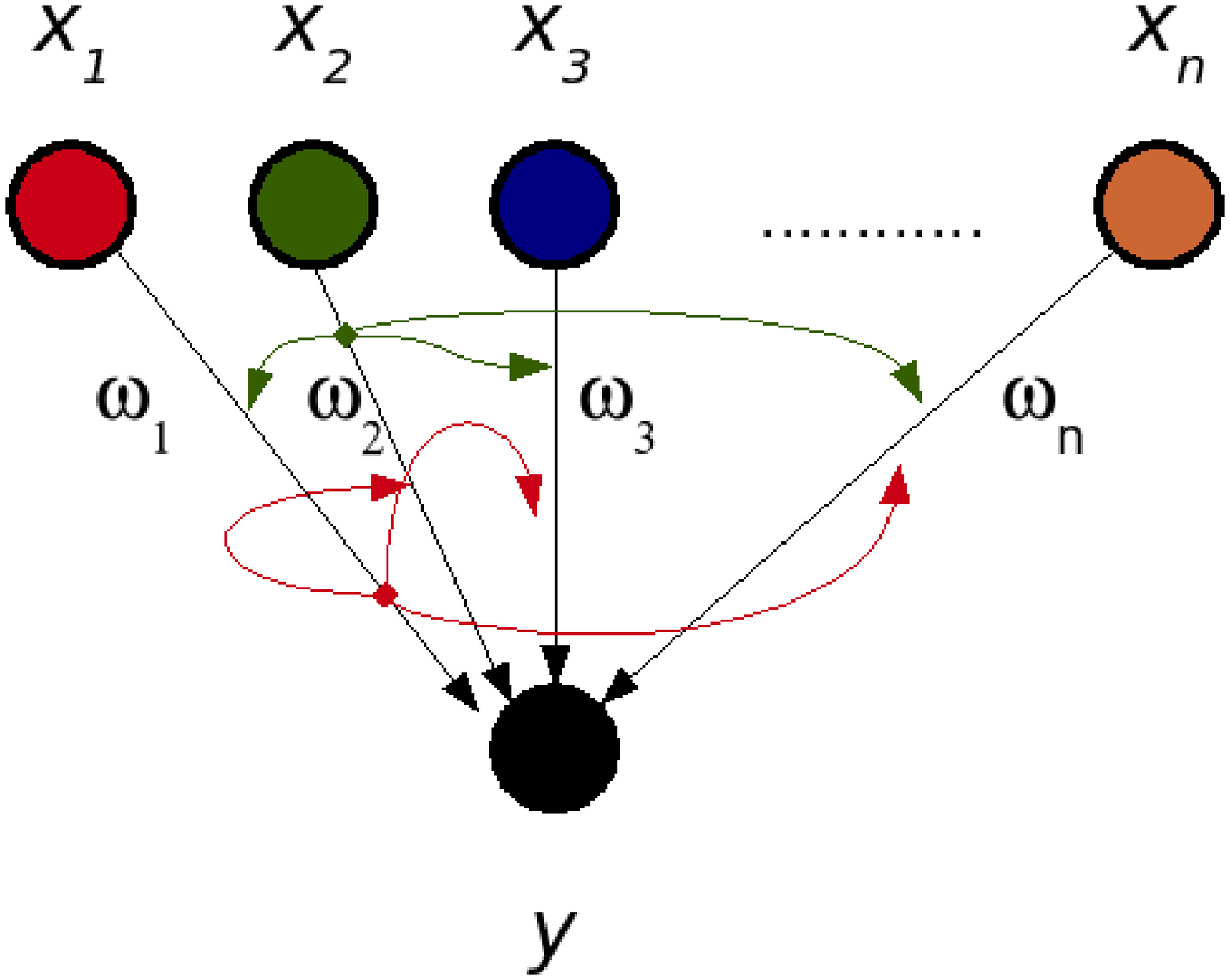}
\end{center}
\caption{\emph{A. Three spines on a short dendritic segment are shown. If Hebbian adjustment occurs at the middle synapse, a factor (red dots) such as calcium, diffuses to nearby synapses and affects Hebbian adjustment there. B. Input neurons (activities $x_i$) converge on an output neuron (output $y$) via weights $\omega_i$. Coincident activity at the synapses comprising a weight (e.g. $\omega_1$ or $\omega_2$) leads to modification of that weight and of other weights. The left diagram shows the case where only the immediate neighboring connections (each made up of one synapse) are affected. The right diagram shows the case where all connections are equal neighbors (either because each has many synapses dispersed randomly over the dendrite, or because synapses move around). The curved red arrows from $\omega_1$ to $\omega_n$ shows that periodic boundary conditions are assumed (i.e., $\omega_1$ affects $\omega_2$ and $\omega_n$ equally).}}
\end{figure}

\section{Methods}

   As a prelude to analyzing the dynamics of inspecific learning, we revisit the Oja original model with zero error, and the methods used to establish its asymptotic behavior~\cite{Botelho}~\cite{Oja1}~\cite{Oja2}~\cite{Herz}. In other words: for a size $n \in \mathbb{N}, \; n \geq 2$, we want to know whether or not a vector ${\bf w} \in \mathbb{R}^{n}$ stabilizes under iterations of the function:\\

 $f \colon \mathbb{R}^{n} \, \to \, \mathbb{R}^{n}, \quad f({\bf w})={\bf w}+\gamma[{\bf Cw}-({\bf w}^{T}{\bf Cw}){\bf w}]$\\

\noindent A vector ${\bf w} \in \mathbb{R}^n$ is a fixed point for $f$ if\\

  $f({\bf w})={\bf w}+\gamma[{\bf Cw}-({\bf w}^{T}{\bf Cw}){\bf w}]={\bf w} \; \Leftrightarrow \; {\bf Cw}=({\bf w}^{T}{\bf Cw}){\bf w}$\\

\noindent An equivalent set of conditions is:

    \[  \left\{ \begin{array}{c}
          \mbox{${\bf Cw}=\lambda_{\bf w} {\bf w}$} \\
          \mbox{$\lambda_{\bf w}={\bf w}^{T}{\bf Cw}$}
          \end{array}
       \right.\Leftrightarrow \left\{ \begin{array}{c}
          \mbox{${\bf Cw}=\lambda_{\bf w} {\bf w}$} \\
          \mbox{$\lambda_{\bf w}=\lambda_{\bf w} {\bf w}^{T}{\bf w}$}
          \end{array}
       \right. \]

\noindent These conditions translate as: ``${\bf w}$ is an eigenvector of
${\bf C}$''. In case ${\bf C}$ is invertible (i.e. all its eigenvalues are nonzero), ${\bf w}$ is a unit eigenvector of ${\bf C}$ with the Euclidean norm.

  Consider an orthonormal basis ${\cal{B}}$ of eigenvectors of ${\bf C}$
  (with respect to the Euclidean norm on $\mathbb{R}^n$). An
  eigenvector ${\bf w} \in {\cal{B}}$ with eigenvalue $\lambda_{\bf w}$ is
  a hyperbolic attractor for $f$ if all eigenvalues of the $n \times n$ Jacobian matrix $\displaystyle{(Df_{\bf w})_{ij}=\left( \frac{\partial f_i}{\partial w_j}({\bf w}) \right)}$ are less than one in absolute value. We calculate $Df_{\bf w}$ and find that ${\cal{B}}$ is also a basis of eigenvectors for $Df_{\bf w}$ (see Appendix B). The corresponding eigenvalues are $1 - 2 \gamma \lambda_{\bf w}$ (for the eigenvector ${\bf w}$) and $1 - \gamma(\lambda_{\bf w} - \lambda_{\bf v})$ (for any other eigenvector ${\bf v} \in {\cal{B}}, \, , \, {\bf v} \not= {\bf w}$). Therefore, a set of equivalent conditions for ${\bf w}$ to be a hyperbolic attractor for $f$ is:\\

  $ \lvert 1 - \gamma(\lambda_{\bf w} - \lambda_{\bf v}) \rvert < 1 \, , \text{ for all } {\bf v} \in {\cal{B}} \, , \, {\bf v} \not= {\bf w}$\\

$ \lvert 1 - 2 \gamma \lambda_{\bf w} \rvert < 1$\\

\noindent
So  ${\bf w}$ is a hyperbolic fixed point of {\em f} if and only if:\\

$(i) \; \; \lambda_{\bf w} > \lambda_{\bf v} \, , \, \text{ for all }\, {\bf v} \not= {\bf w}$ \,
(i.e. $\lambda_{\bf w}$ is the maximal eigenvalue)\\

$(ii) \; \; \gamma < \frac{1}{\lambda_{\bf w}}$ (in particular $\gamma < \frac{2}{\lambda_{\bf w} - \lambda_{\bf v}} \, , \, \text{ for all }{\bf v} \not= {\bf w}$) \\

These conditions are always satisfied provided: \emph{(i)} {\bf C} has a maximal eigenvalue of multiplicity one and \emph{(ii)} $\gamma$ is small enough ($\gamma < \frac{1}{\lambda_{\bf w}}$).\\

In conclusion: under conditions $(i)$ and $(ii)$, the network learns the first principal component (PC1) of the distribution ${\cal P}({\bf x})$. The learning of the principal component requires a relationship between the rate of learning $\gamma$ and the input distribution
${\cal P}({\bf x})$: if the maximal eigenvalue of the correlation
matrix ${\bf C}$ is large (i.e. if the variance of the input patterns' projections on PC1 is high), the network has to learn slowly in order to achieve convergence. Moreover, the convergence time along each eigendirection is given by the inverse of the magnitude of the corresponding eigenvalue of $Df_{\bf w}$ (see the simulations in Figure 2).\\

To formalize learning inspecificity we introduced an error matrix $E \in {\cal{M}}_{n}(\mathbb{R})$ that has positive entries, is symmetric and equal to the identity matrix $I \in {\cal{M}}_{n}(\mathbb{R})$ when the error is zero. We studied the asymptotic behavior of the new system, using the
approach outlined in Section 2 (also see Appendix A). The inspecific learning
iteration function becomes:

  $$ f^{\bf E}({\bf w})={\bf w}+\gamma[{\bf ECw}-({\bf w}^{T}{\bf Cw}){\bf w}]$$

\noindent Here also, ${\bf w}$ is a fixed point of $f^{\bf E}$ if and only if
it is an eigenvector of ${\bf EC}$ with eigenvalue
$\lambda_{\bf w}=({\bf w}^{T}{\bf Cw}){\bf w}>0$. Furthermore, ${\bf w}$ is a hyperbolic attractor of $f^{\bf E}$ if and only if $\lambda_{\bf w}$ is the
principal eigenvalue of ${\bf EC}$ and $\gamma<\frac{1}{\lambda_{\bf w}}$.

The error-free rule maximizes the variance of the output neuron $\lambda_{\bf w}$ and therefore, with Gaussian inputs, also maximizes the mutual information between inputs and outputs (see Discussion). Altough the erroneous rule no longer maximizes the output variance, it tolerates a faster learning rate. Conversely, at a fixed $\gamma$, learning is slowed by error.

\subsection{The error matrix}

One way in which an incorrect strengthening of a silent synapse can occur is by diffusion of a messenger such as calcium from one spine head to another, as illustrated in Figure 1a.

If we assume that the output neuron is connected (at least potentially) to all the input neurons~\cite{Stepanyants} then the amount of error
depends on the number of synapses each input neuron makes with the
output neuron (relative to the dendritic length $L$) as well as factors such as the space constant for dendritic calcium diffusion $\lambda_c$~\cite{Zador}, the Hill coefficient for calcium action $h$~\cite{DeKoninck}~\cite{Lisman2}, and the amount of head/shaft/head calcium attenuation $a$. We can define a per ``synapse error factor'' $b \in [0,1]$.

\begin{equation}
b \sim \frac{a^{h} \lambda_{c}}{L}
\end{equation}

 \noindent or equivalently a ``synaptic quality" $q \in [0,1]$,
 $q=1-b$ (see Text S1 for definitions and details). This formula says that the per synapse error $b$ is proportional to two factors: the  ratio of the length constant for calcium spread $\lambda_c$ to the dendritic length $L$ and the effective calcium coupling constant $a^h$ between two adjoining spines. It assumes that as extra inputs are added, the dendritic length remains the same (see Discussion).

The probability $Q$ of the correct synapse being strengthened
depends on $b$ and on the network size $n$. In Text S2 we analyze a plausible model and we develop two approximations for $Q=Q(n,b)$.\\

   $\bullet$ the \textbf{continuous model}, where weights adjust continuously and $Q=\frac{1}{nb+1}$. \\

   $\bullet$ the \textbf{discrete model}, where weights adjust discretely and $Q=(1-b)^{n}$. \\

\subsection{Error spread}

   We now consider different possibilities for the way that the part of the Hebbian update $x_i y$ could spread to different connections, presumably as a result of inttracellular diffusion of messengers such as calcium. In general this will reflect the particular anatomical relationships between synapses, expressed by \mbox{\boldmath ${\cal E}$}, which could change as learning proceeds. We examine two extreme cases. First, each connection is made of a single fixed synapse (e.g., a parallel fiber-Purkinje cell connection~\cite{Llinas}). In the second case, all connections are equivalent (``tabula rasa''~\cite{LeBe}~\cite{Elman}).

1.  The ``{\bf nearest neighbor}'' model: Each connection consists
of a single fixed synapse, and calcium only spreads to two
nearest neighbor synapses. ${\bf E}$ then has diagonal
elements $Q$ and off diagonal elements $\frac{1-Q}{2}$.

$$ {\bf E}=   \left( \begin{array}{cccccc}
       Q&\epsilon&0&\cdot&\cdot&\epsilon \\
       \epsilon&Q&\epsilon&0&\cdot&0 \\
       0&\epsilon&Q&\epsilon&\cdot&0 \\
       \cdot&\cdot&\cdot&\cdot&\cdot&\cdot \\
       0&\cdot&\cdot&\epsilon&Q&\epsilon \\
       \epsilon&0&\cdot&\cdot&\epsilon&Q
         \end{array} \right) ,$$

\noindent The appearance of $\epsilon$ in the top right and bottom right corners reflects periodic boundary conditions. We can define a ``trivial'' error rate $\epsilon = 1/3$ for which Hebbian adjustments lacks specificity, which is marked in Figure 3a as a red asterisk on each curve.\\

2.  The ``{\bf error-onto-all}'' model: All connections are equally ``distant'' from each other, so
that there are no privileged connections. All offdiagonal elements
of ${\bf E}$ are then equal to $\frac{1-Q}{n-1}$.

$$ {\bf E}=   \left( \begin{array}{cccccc}
       Q&\epsilon&\epsilon&\cdot&\cdot&\epsilon \\
       \epsilon&Q&\epsilon&\epsilon&\cdot&\epsilon \\
       \epsilon&\epsilon&Q&\epsilon&\cdot&\epsilon \\
       \cdot&\cdot&\cdot&\cdot&\cdot&\cdot \\
       \epsilon&\cdot&\cdot&\epsilon&Q&\epsilon \\
       \epsilon&\epsilon&\cdot&\cdot&\epsilon&Q
         \end{array} \right)$$

\noindent It is important to notice that the error matrix in this case becomes singular when $Q=\epsilon$, i.e. when the update leak to each erroneous connection is as large as the update at  the right connection. We call this value the ``trivial error value" $\epsilon_0(n)$, which corresponds to $b_0(n)=1-q_0(n)=1-1/\sqrt[n]{n}$ in the discrete model, and to $b_0(n)=1-q_0(n)=1/n$ in the continuous model. For all biological purposes, we need only consider errors smaller than the trivial value.\\

 This arrangement could arise in two nonexclusive different ways:

a. Each connection is composed of a very large number $\alpha$ of fixed synapses, such that all possible configurations of synapses occur.

b. Synapses do not have fixed locations, but appear and disappear randomly at
all possible locations (i.e. ``touchpoints''~\cite{LeBe}~\cite{Stepanyants} where axons approach the dendrite close enough that a new spine can create a synapse). In this case, assuming the dendrite and axonal geometry are fixed~\cite{Svoboda1}~\cite{Gan}~\cite{Svoboda2}, the postsynaptic neuron has a reservoir of ``potential'' synapses~\cite{Stepanyants}, composed of 2 shifting subsets: anatomically existing synapses and a reservoir of "incipient" [3] synapses where spines could form (see Text S2). In order to maintain constant weights (in the absence of learning), each synapse would have to be replaced by another synapse of
equal strength and, possibly, connectivity. This could be done most simply if only zero-strength (``silent'') synapses appear and disappear (since then connectivity would not have to be conserved). There is evidence this is the case~\cite{Sabatini1}.

If synapses appear and disappear, one has the problem that if all synapses are equally plastic (have the same learning rate $\gamma$), stochastic changes in the overall number of synapses comprising a connection will change the overall learning rate at a connection. (In the simplest case, if a number of new silent plastic synapses happen to appear at a connection, while the overall weight is unchanged, the learning rate will be increased). One way to prevent this would be to ensure that only one of the synapses comprising a connection is plastic~\cite{Adams1} but this is a nonlocal rule. Another way would be for the average number of potential synapses comprising a connection to be reasonably high (perhaps $\sim 50$ ) so that fluctuations are relatively small. In several cases the average number of actual synapses at a connection is around 5~\cite{Markram2}~\cite{Barbour} and since these may only form about 10 $\%$ of the total (potential) synapses, learning rates at different connections wold be fairly similar (and of course identical when time-averaged). Of course for this rough-and ready solution to work, axons and dendrites would have to intersect sufficiently often, implying a high degree of branching. Although there have been some claims that weak synapses are more plastic~\cite{Matsuzaki}, other evidence suggests that all synapses are equally plastic~\cite{Kopec}; this could be achieved if strengthening added a new plastic ``unit'' to each synapse, with previously added units all rendered implastic~\cite{Adams1}~\cite{Lisman3}(also see Discussion). A related issue is that the synapses comprising a connection will be at different electrotonic distances along the dendrite, and therefore will influence spiking differently, and have different effective learning rates. Rumsey et al.~\cite{Rumsey} have proposed that a separate antiSTDP can be used to equalize efficacy of weights.

   There is strong evidence for both these ways to achieve complete connectivity~\cite{LeBe}~\cite{Sabatini1}, and we think this is the most biologically plausible assumption, and it will be the principal target of our analysis.\\

The stabilized weight vector of the modified (inaccurate) Oja model differs from the principal component of ${\bf C}$. The analysis in Section 3 and Text S1 shows that the inspecific learning algorithm should still converge, but now to the principal eigenvector ${\bf w}_{\bf EC}$ of ${\bf EC}$ rather than to the principal component ${\bf w_C}$ of the input distribution. Since the output of the Oja neuron allows optimal input reconstruction (at least in the least squares sense; see Discussion), Hebbian infidelity leads
  to suboptimal performance.  We quantified the effect of infidelity as the
  cosine of the angle $\theta$ between the principal component ${\bf w_C}$  of ${\bf C}$ and the principal eigenvector ${\bf w_{EC}}$ of ${\bf EC}$, which are the stabilized weight vectors in absence and respectively presence of error:

$$\cos(\theta)=\frac{ {\bf w_{C}}^T{\bf w_{EC}}}{\| {\bf w_{C}} \| \cdot \| {\bf w_{EC}} \|}$$

   We examined how this measure of error depends on parameters such as the size $n$ and the input error $\epsilon$ for a given ${\bf C}$. As the analysis for arbitrary input distributions is rather intractable, we detail only a few simple cases of uncorrelated (Section 4.1) and correlated inputs (Section 4.2), illustrating the results with Matlab plots and simulations.

\section{Results}

\begin{figure}[h!]
\includegraphics[scale=0.4]{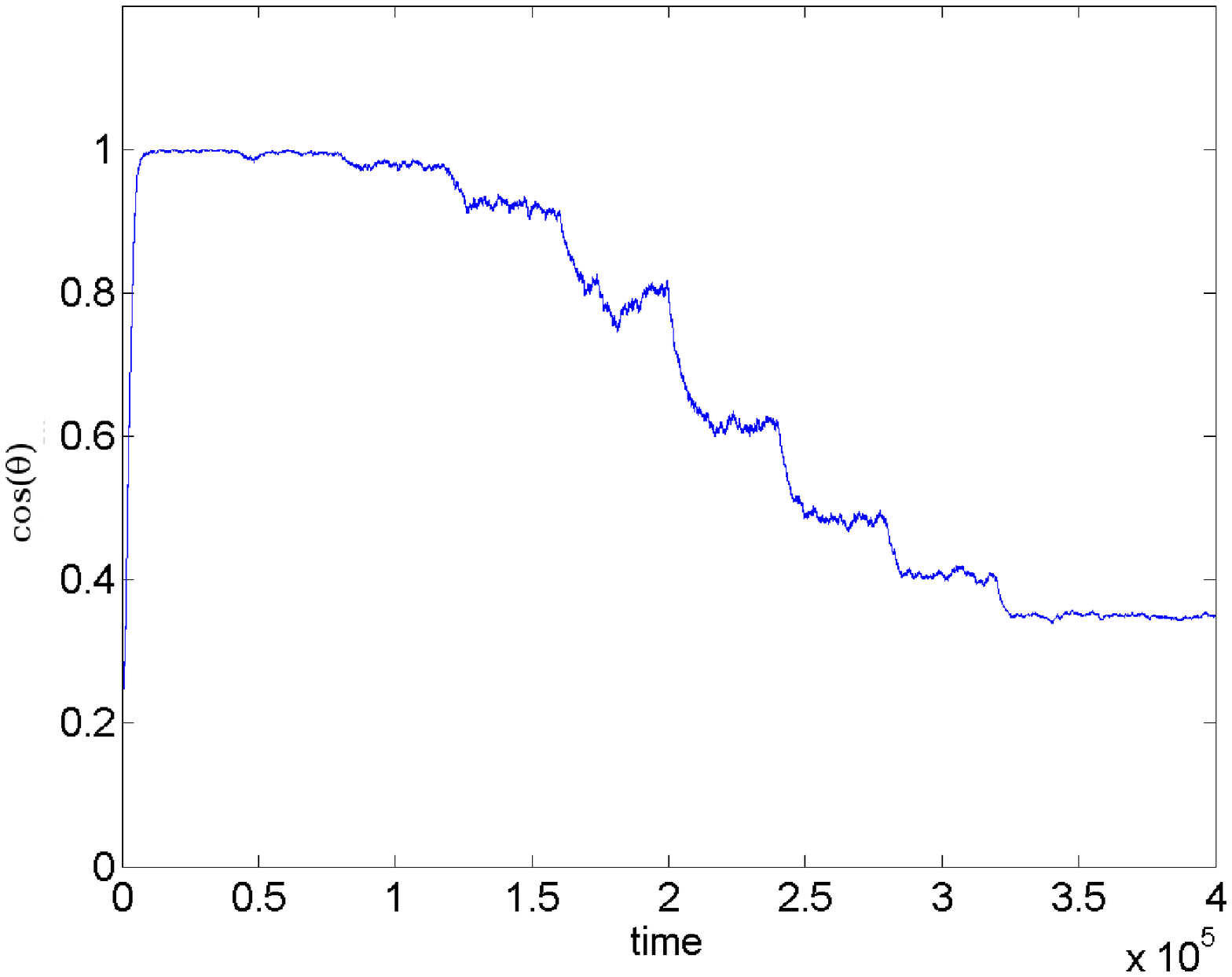}
\includegraphics[scale=0.4]{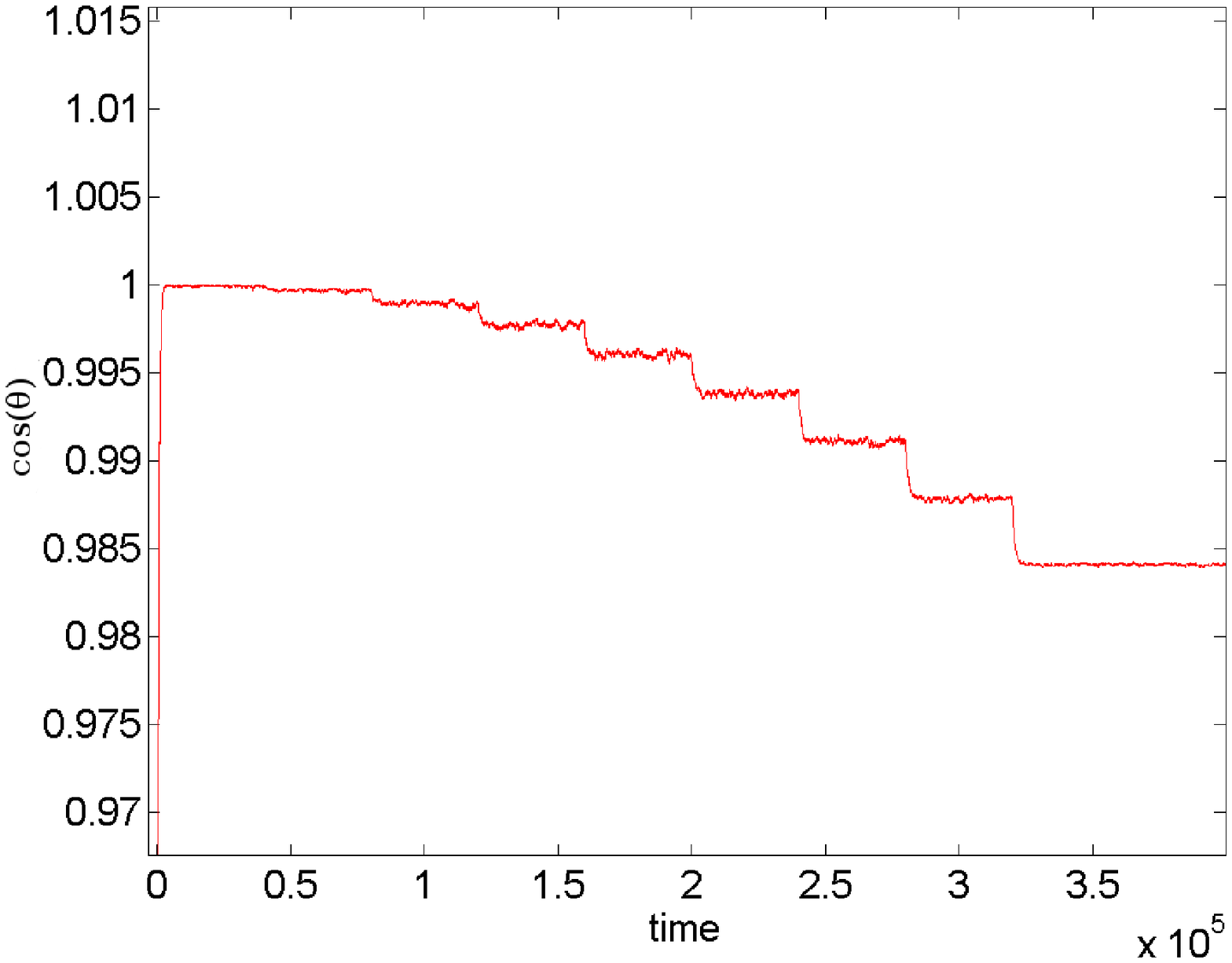}
\caption{\emph{Effect of errors on performance in the Oja network for $n=10$ neurons. The plots represent the cosine of the angle $\theta$ between the weight vector and the principal component, at each time-step of the updating process. Left: The inputs are uncorrelated gaussian vectors with one of the sources at a variance of 2 compared to the others which are all 1. Right: Correlated inputs were generated by mixing sources with equal variance with a random mixing matrix, with elements distributed uniformly between 0 and 1. In both cases, the total error {\Large{$\epsilon$}}=$1-Q=(n-1)\epsilon$ was initially set to zero and the weight vector converged very quickly to the first PC ($\cos(\theta)=1$) and remained there with minor fluctuations. The error was then increased from zero to $0.8$ in steps of $0.1$ each $4 \times 10^4$ epochs, producing approximately stepwise decreases in performance. The equilibration time increased as error increased; the step heights and the associated fluctuation increased and then decreased. Note that in the correlated case error produces only small decreases in performance, since the principal component already points approximately in the direction $(1,1,...1)$.}}
\end{figure}

We start with examples of simulations of the behavior of the erroneous rule in the error-onto-all case, using either uncorrelated (Figure 2a) or correlated (Figure 2b,c) inputs. In all cases the network is initialised with random weights. In the uncorrelated case and in the absence of error, the correct principal component is learned rapidly and accurately. The small fluctuations away from PC1 reflect the nonzero learning rate; they are most obvious at error rates for which the dependence of performance on control parameters is steepest (see Figure 5a). Figures 2b and 2c show that the effect of error on performance depends on the degree of correlation present. In all cases, performance (measured by $\cos(\theta)$) gradually deteriorates with progressive increase in error, although the magnitude of the decrease depends on error and on correlation. The remainder of our results explore these effects in more detail, using calculations and implicit analysis. Figure 2 also shows that learning is somewhat slowed by inaccuracy, as expected; however, we do not analyze learning kinetics further here.

\subsection{Uncorrelated inputs}
\noindent

 This section shows how network performance depends on the quality factor $q \in [0,q_0(n)]$ (or alternatively on the error factor $b=1-q$) in the case of uncorrelated inputs. We illustrate this dependence by a combination of Matlab plots and analytical results.

For the uncorrelated inputs case, we consider a diagonal ${\bf C}$ with
higher variance on the first component:

\begin{eqnarray}
 {\bf C}=\left( \begin{array}{ccccc}
       \lambda&0&0&\cdot&0 \\
       0&1&0&\cdot&0 \\
       0&0&\cdot&\cdot&\cdot \\
       \cdot&\cdot&\cdot&1&0 \\
       0&0&\cdot&0&1
       \end{array} \right)
 \end{eqnarray}
\noindent where $\lambda>1$, so that ${\bf w_{C}}=(1,0,...,0)^{T}$ . In this case,

$$\cos(\theta)=\frac{\lvert ({\bf w_{EC}})_{1} \rvert}{\| {\bf w_{EC}} \|}$$
\noindent is our measure of the system's performance.

   We studied how $\cos(\theta)$ changes with the error (either $\epsilon$ or $b$), $n$ and $\lambda$. We numerically calculated $\cos(\theta)$ as a function of the error in the two cases where the error is apportioned to two neighbors (nearest-neighbor model, Figure 3a) or to all other connections (error onto all model, Figure 4a).

\begin{figure}[h!]
\includegraphics[scale=0.50]{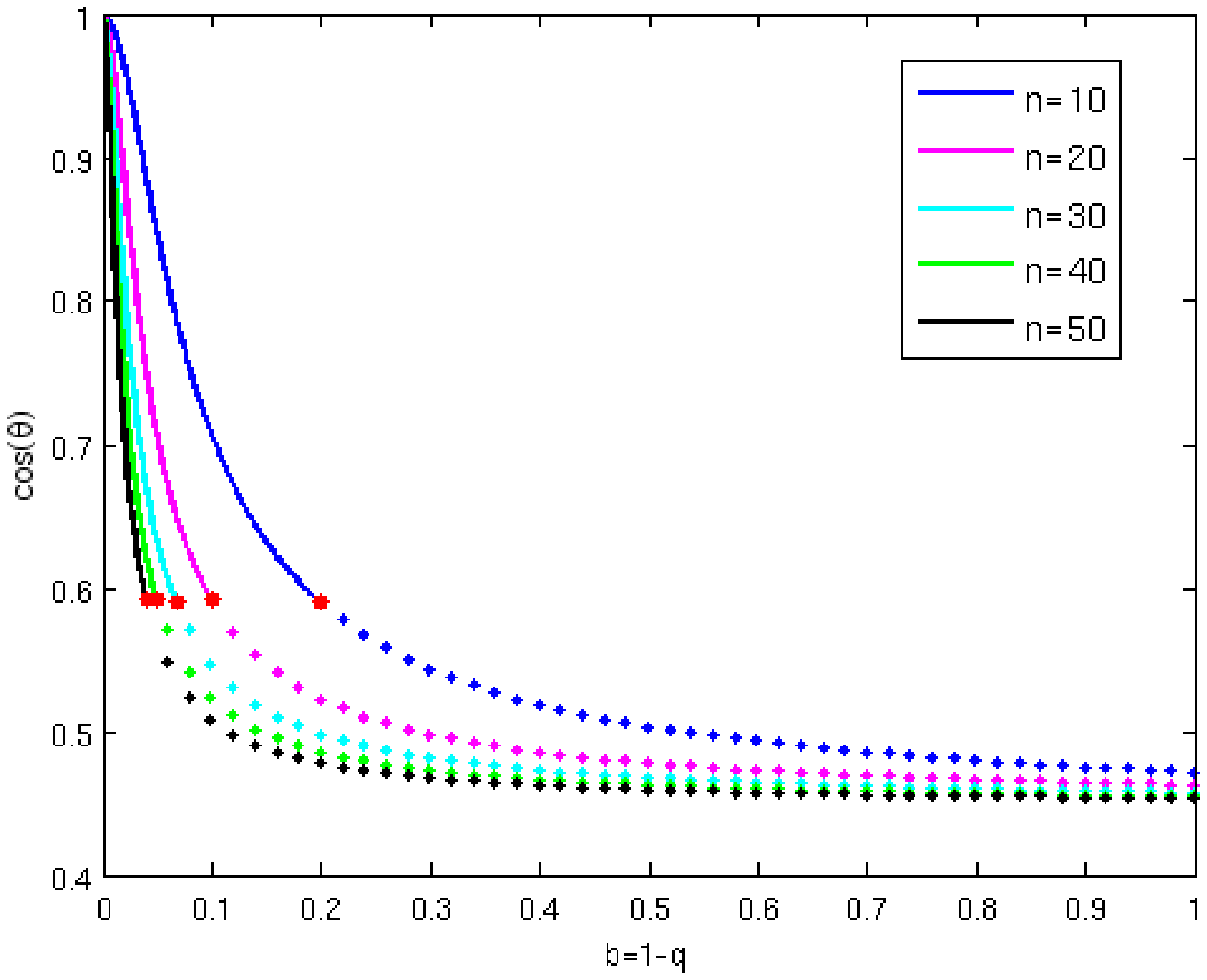}
\includegraphics[scale=0.5]{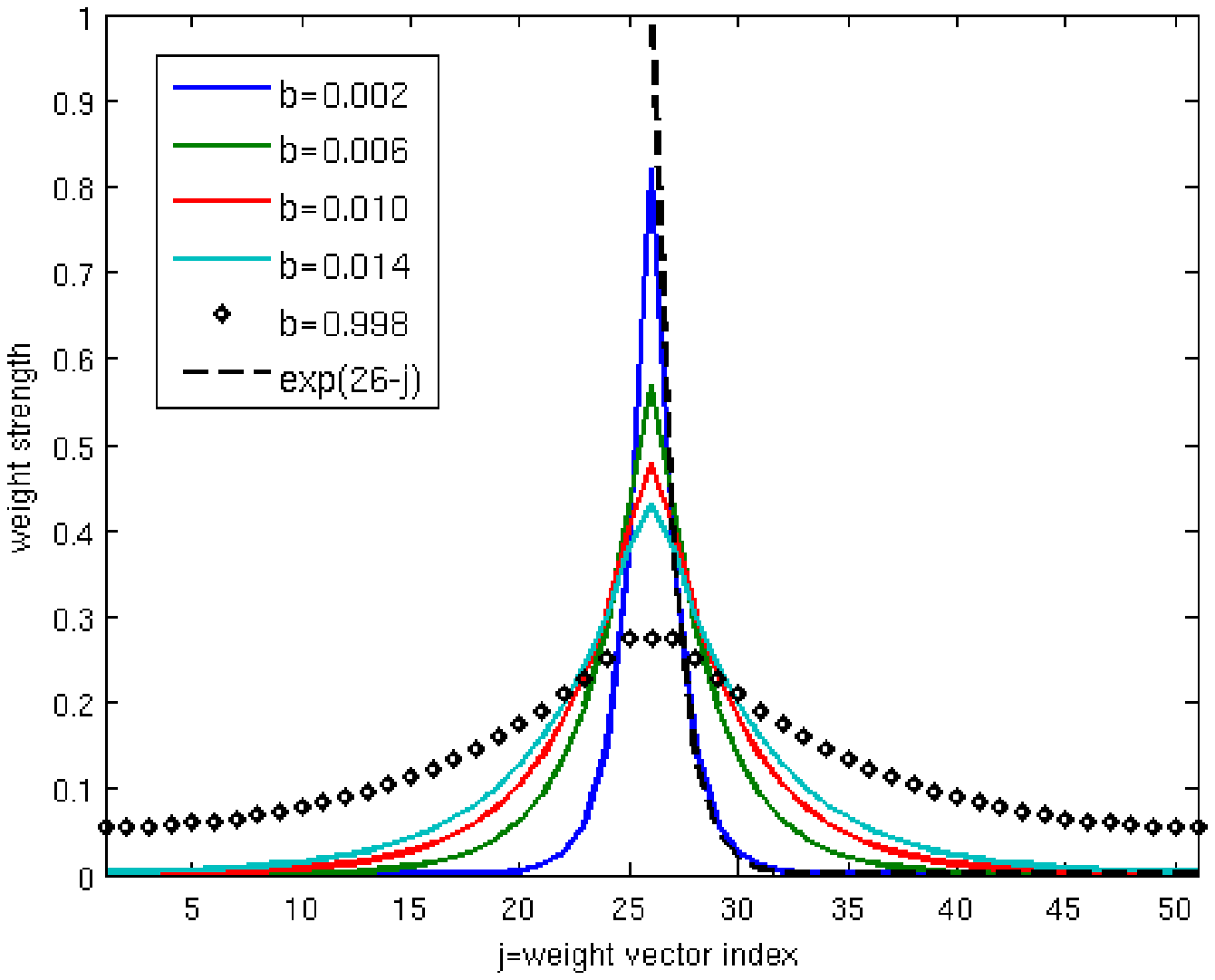}
\caption{\emph{Dependence of $\cos (\theta)$ on the error factor $b$ in the case of uncorrelated inputs with $\lambda=2$ for the continuous error, nearest-neighbour model. Each curve corresponds to a different $n$, as shown in the legends. Left: Continuous error, nearest neighbor model. Note that for increasing $n$ values, the value of the total error {\Large{$\epsilon$}} increases at any fixed per synapse error $b$, reducing performance ($\cos(\theta)$). The curves are shown as solid lines up to the trivial error value where $Q=\epsilon=1/3$ (i.e., $b=2/n$) at which learning is inspecific. (red dots); beyond this point the curves are unbiological and are shown dotted. Right: The distribution of weights of the asymptotically stable weight vector (the principal eigenvector of ${\bf EC}$), for fixed network size $n=51$, and fixed variance $\lambda=1.1$, but different values of the per synapse error $b$. The lower the quality, the more similar the weights become. All the $b$ values shown are less than trivial, except for the curve marked with diamonds, where almost all the updates are transferred to neighbors. With the exception of the weight on the high-variance neuron labeled $\#$26, the weights decay approximately exponentially as a function of distance from the high-variance neuron. This is illustrated by the black dashed curve, which is a shifted exponential with space constant of one unit (neuron). The space constant calculated from Equation 7 in~\cite{Adams1} for the corresponding values of $b$, $n$ and $\lambda$ is $0.7$ neurons.}}
\end{figure}

The curves in the nearest-neighbor model (Figure 3a) can be understood in the following way. First consider a curve at a given network size. As outlined in the Methods, in the absence of error the high variance connection grows more rapidly than the low variance connections, eventually completely winning, so the final weight vector points in that direction. However, in the presence of error, the immediate neighbors strengthen more than they would have in the absence of error, as a result of leakage from the preferred connection (see Figure 3b); this means that future patterns will produce extra strengthening of those neighboring connections (because they are stronger and so produce larger outputs); this extra strengthening at the neighbors leads to increased strengthening of the neighbors of the neighbors, and so on down the line. Since the weight vector is normalised, these ``wrong'' strengthenings combine to reduce the preferred weight, although as long as learning shows some specificity, the preferred final weight is always strongest (see Figure 3b).

Figure 3b shows the distribution of equilibrium weights as a function of ``distance'' from the preferred neuron and of error $b$. The nearest neighbor case corresponds to the ``fitness'' model we simulated in previous work, and analysed in the large $n$ limit~\cite{Adams1}, with ``fitnesses'' being the input variances. In that model, the weight distribution on the nonpreferred connections followed a double exponential function of distance. For sufficiently small error (or large $\lambda$), the distribution of weights on the nonpreferred (low variance) connections is close to the single exponential distribution found in this limit in the ``fitness'' model (see dashed curve in Figure 3b); we also found that the ``space constant'' for the weight  distribution varied in the expected manner with the Hebbian error (Figure 3b) and with the variance $\lambda$.

\begin{figure}[h!]
\includegraphics[scale=0.5]{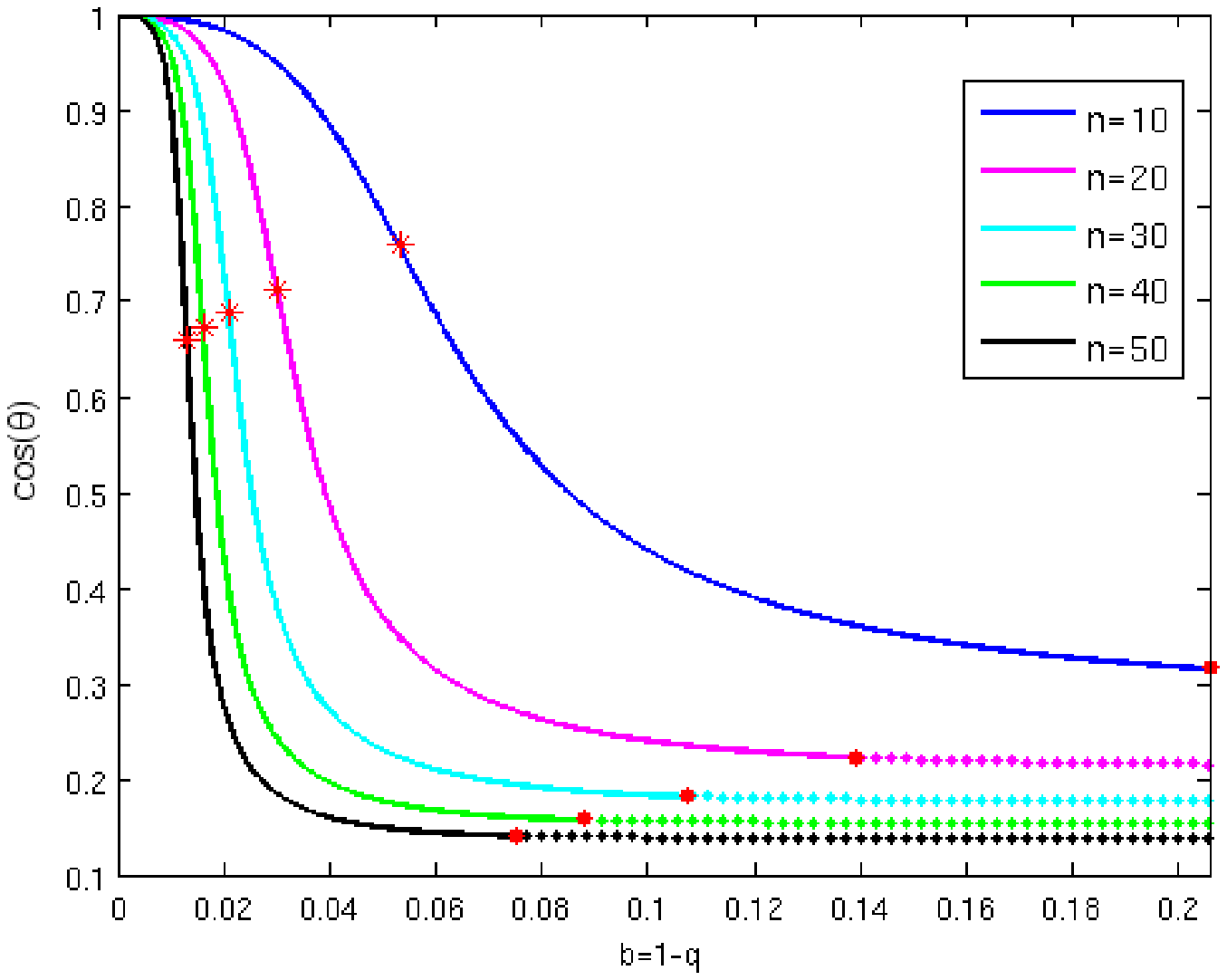}
\includegraphics[scale=0.5]{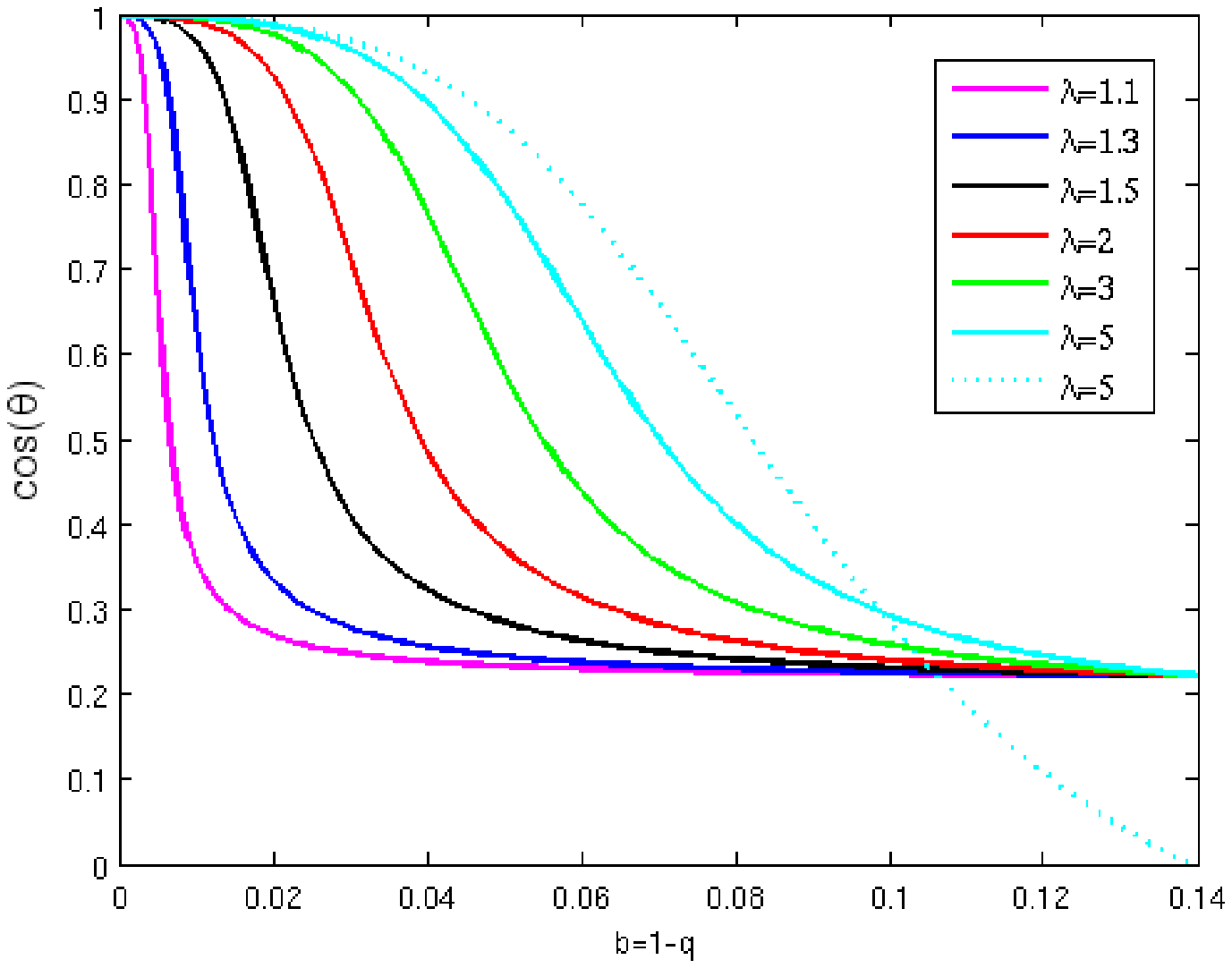}
\caption{\emph{{\bf Error-onto-all, discrete updates model.} Left: Dependence of $\cos (\theta)$ on the error factor $b$ in the case of uncorrelated inputs with $\lambda=2$. The performance is measured as $\cos(\theta)$, where $\theta$ is the angle between the principal eigenvector of ${\bf EC}$ and the principal component of ${\bf C}$. Each color-coded curve corresponds to a different network size $n$, as shown in the legend. The curves were plotted as solid lines for $b$ between zero and the trivial value $b_0(n)=1/\sqrt[n]{n}$, and as dotted lines for the error $b$ larger than the trivial value, because this range is not biological. The point of steepest downward slope (inflection point) is marked on each graph by a red asterisk. Consistently with our calculations, the inflection point is always situated between zero and the trivial value $b_0(n)$, getting arbitrarily close to zero for large enough $n$. Note that the $n=10$ curve agrees almost exactly with the results in Figure 2a, if the $b$ values are converted to the corresponding {\Large{$\epsilon$}} values. Right: For a fixed network size $n=20$ and different values of the input variance $\lambda$, we show the dependence of the output performance $\cos(\theta)$ on the synaptic error $b$, for $b \in [0,b_0(n)]$. Each curve corresponds to a different $\lambda$ value from $\lambda=1.1$ to $\lambda=5$. When the variance is very close to $\lambda \to 1$, the stable weight vector approaches ${\bf w}=(1,1....1)$ independently of the error, so $\cos(\theta) \to 1/\sqrt{n} \sim 0.22$ for all values of $b \neq 0$. Also, $\cos(\theta)=1/\sqrt{n}$ for all $\lambda$ at $b=b_0(n)$. The performance improves with larger variance, which agrees with our analytical results. The dotted line shows the perturbation approximation for $\lambda = 5$ (see Text), which works well only at low error.}}
\end{figure}

In the remainder of the paper we focus on the error-onto-all model, in which the quality of the network is $Q>0$ and the error is $\epsilon=\frac{1-Q}{n-1}$. We will present in detail only the discrete update model, since this seems to be more biologically realistic~\cite{Petersen}~\cite{OConnor} ~\cite{Bagal}; the continuous case is rather similar and treated in Text S2. Numerical calculations of the performance at different per synapse error values at various network sizes for the uncorrelated case are shown in Figure 4a. The curve for $n = 10$ is plotted up to the trivial error value $b=1/\sqrt[10]{10} \sim 0.205$ where learning is completely inspecific.  There is a smoothly increasing degradation of performance with error, which drops to a much lower value for inspecific learning than seen in the previous cases, since error affects all nonpreferred inputs equally (for $b = b_0$, the weight vector is parallel to $(1,1,...1)^T$, so the limiting cosine for the case $n=10$ is $1/\sqrt{10}=0.316$). In the remaining plots in Figure 4a, the unbiological points of the curves beyond the trivial value are shown dotted.

Figure 4b shows plots of performance against $b$ for different values of variance $\lambda$, all for the case $n = 20$. There is a very large change in performance for small increases of $\lambda$ above one especially at low error values. For $\lambda = 1$, all eigenvalues are equal and the corresponding eigenvectors are only half-stable. A tiny bit of error stabilises the behavior, so only the preferred weight is selected, although never perfectly.

We obtained an approximation for $\cos(\theta)$ at small $\epsilon$ values using perturbation theory ~\cite{Kahn}:

$$\cos(\theta)=\frac{\lambda-1}{(\lambda-1)^2+(n-1)\lambda^2 \epsilon^2 / (1-n \epsilon)^2} $$

Figure 4b shows that this formula agrees well with the exact results at sufficiently small $\epsilon$.

We now proceed to an analytic treatment of these numerical results. The characteristic polynomial of the error matrix is:

   $$P^{\bf E}(x)=det({\bf E}-x{\bf I})=(Q-\epsilon-x)^{(n-1)}(1-x)$$

\noindent Note that ${\bf E}$ is invertible, except when $Q=\epsilon$, and
$Q$ and $\epsilon$ themselves depend on biological parameters (see Text 2).

The maximal eigenvalue $\mu$ of ${\bf EC}$ is given in this case by the larger solution of the quadratic equation:

$$\mu^2-\mu[\lambda+1+\epsilon(\lambda-1-n \lambda)]+\lambda-n
\lambda \epsilon=0$$

The maximal eigenvector will be in the direction $(s,1,...1)^T$, where $s=s(\epsilon, n, \lambda)$ is a ``selectivity'' value which expresses how strongly one of the weights is favored because one input is more active. This outcome reflects the fact that no weight except that corresponding to the high-variance input is preferred (there are no privileged neighbor relations), so the behavior boils down to competition between the preferred weight and the set of nonpreferred weights, leading to the quadratic equation.

We usually estimate the output performance as $\cos(\theta)$, but here it's simplest to calculate $\tan(\theta)$, which is related to $\cos(\theta)$ by: $\cos(\theta)=1/\sqrt{\tan^2(\theta)+1}$, for $\theta \in [0,\pi/2]$.\\

$\displaystyle{h(q)=\tan(\theta(q))=-\frac{\lambda-1}{\sqrt{n-1}}\frac{\lambda-\mu(q)}{\mu(q)-1}}=\frac{s}{\sqrt{1-s^2}}$\\

$h'=\displaystyle{-\frac{\lambda-1}{\sqrt{n-1}}\frac{\mu'}{(\mu-1)^2}}$\\

$h''=\displaystyle{-\frac{\lambda-1}{\sqrt{n-1}}\frac{\mu''(\mu-1)-2(\mu')^2}{(\mu-1)^3}}$\\

\noindent where all derivatives are with respect to $q$. As
$\mu'>0$, we have $h'<0$ for all $q$. This is consistent with our
simulations: performance decays as the quality factor decreases.

Both the discrete and continuous update models show similar features (Section 3.1 and Text S3). The angle $\theta=\theta(q)$ (measured by its tangent $h(q)=\tan(\theta(q))$) decreases as $q$ goes from $0$ to $1$. In both cases $h(1)=0$, which corresponds to perfect performance for
perfect quality. Also, $h(0) \to 0$ as $n \to \infty$, which
shows that the output degrades more severely with error for
larger values of the network size (because of synaptic ``crowding'').
Moreover, $h'(1) \to \infty$ as $n \to \infty$, which shows that the rate of the angle decay at $q=1$ gets very steep with large $n$. Since the slope is always finite at finite epsilon, there is no ``error catastrophe'' (see Discussion).

A less obvious observation concerns the inflection point on each
graph, where the decay rate (or ``error sensitivity'') of the performance is steepest (see red asterisks in Figures 4a and 5). Although an exact estimate
is intractable, we obtained, using the above expressions for the derivatives of $\tan(\theta)$, a lower bound: the inflection point is
always situated in the interval $[q_0(n),1]$ (or equivalently in $[0,b_0(n)]$, when reporting to synaptic error); see Text S3. Figure 5a further suggests that the inflection point always moves to the left in step with the leftward shift in the trivial error value as $n$ gets larger.

In summary, in the uncorrelated case, high per-synapse quality ensures excellent performance except when inputs are numerous (high $n$), or almost indistinguishable (low $\lambda$). Conversely, since performance only improves very slightly when error is further reduced from initially very low values, it would be difficult for evolution to attain very low error rates. We next asked if these features remain true for correlated inputs.

\subsection{Correlated inputs}

We now study the equilibrium behavior of the network in response to two simple cases of correlated inputs, in the error-onto-all model, with the following covariance matrices:

\begin{eqnarray}
{\bf C}=\left( \begin{array}{ccccc}
       1&\lambda&\xi&\cdot&\xi \\
       \lambda&1&\xi&\cdot&\xi \\
       \xi&\xi&1&\cdot&\cdot\\
       \cdot&\cdot&\cdot&\cdot&\xi \\
       \xi&\xi&\cdot&\xi&1
       \end{array} \right)
\end{eqnarray}

\noindent where $1>\lambda>\xi>0$ (${\bf C}$ has higher covariance on one pair) and

\begin{eqnarray}
{\bf C}=\left( \begin{array}{ccccc}
       \lambda&\xi&\xi&\cdot&\xi\\
       \xi&1&\xi&\cdot&\xi\\
       \xi&\xi&1&\cdot&\cdot\\
       \cdot&\cdot&\cdot&\cdot&\xi\\
       \xi&\xi&\cdot&\xi&1
       \end{array} \right)
\end{eqnarray}

\noindent where $\lambda>1>\xi>0$ (${\bf C}$ has small uniform background covariance with one high variance input).

\begin{figure}[h!]
\includegraphics[scale=0.5]{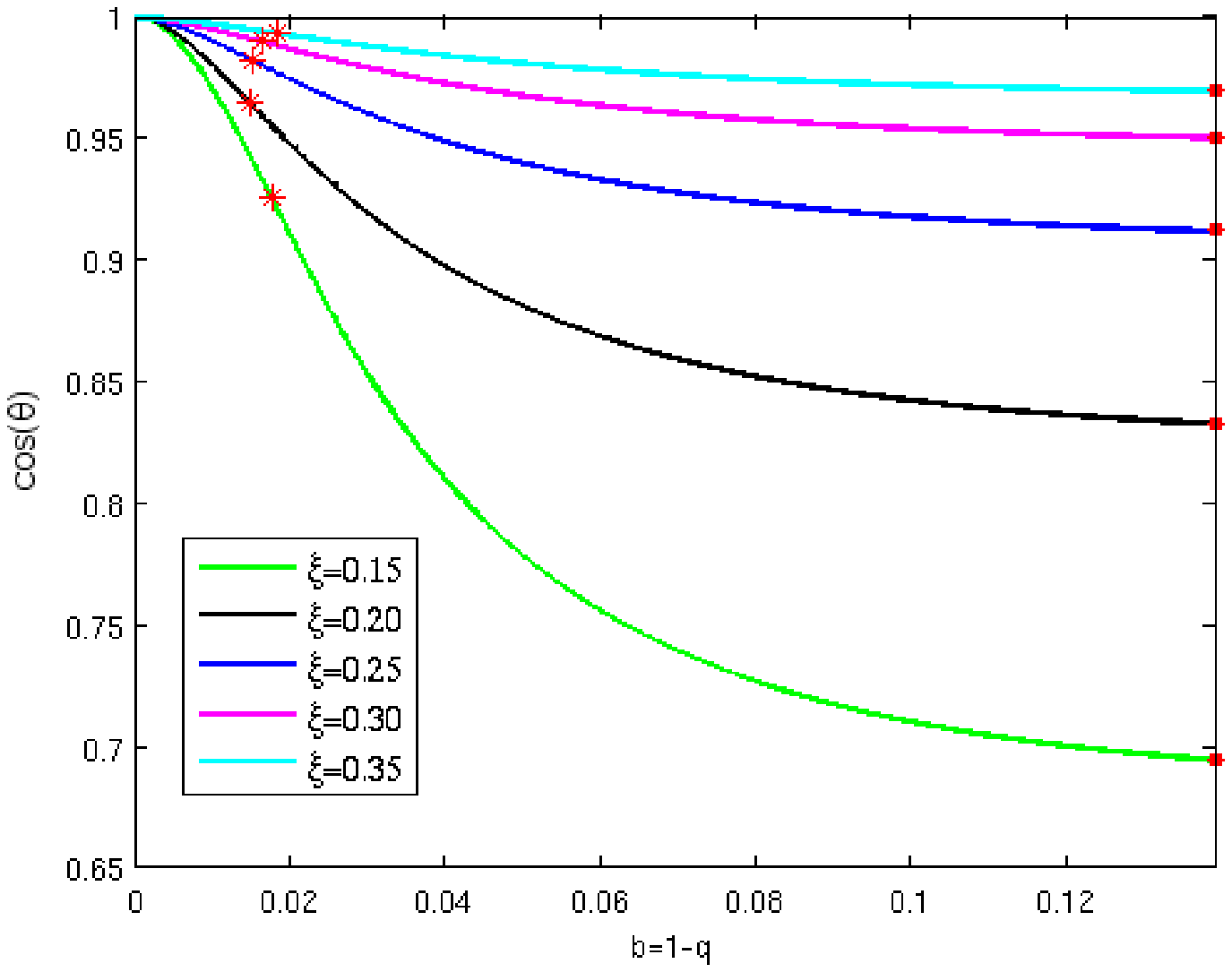}
\includegraphics[scale=0.5]{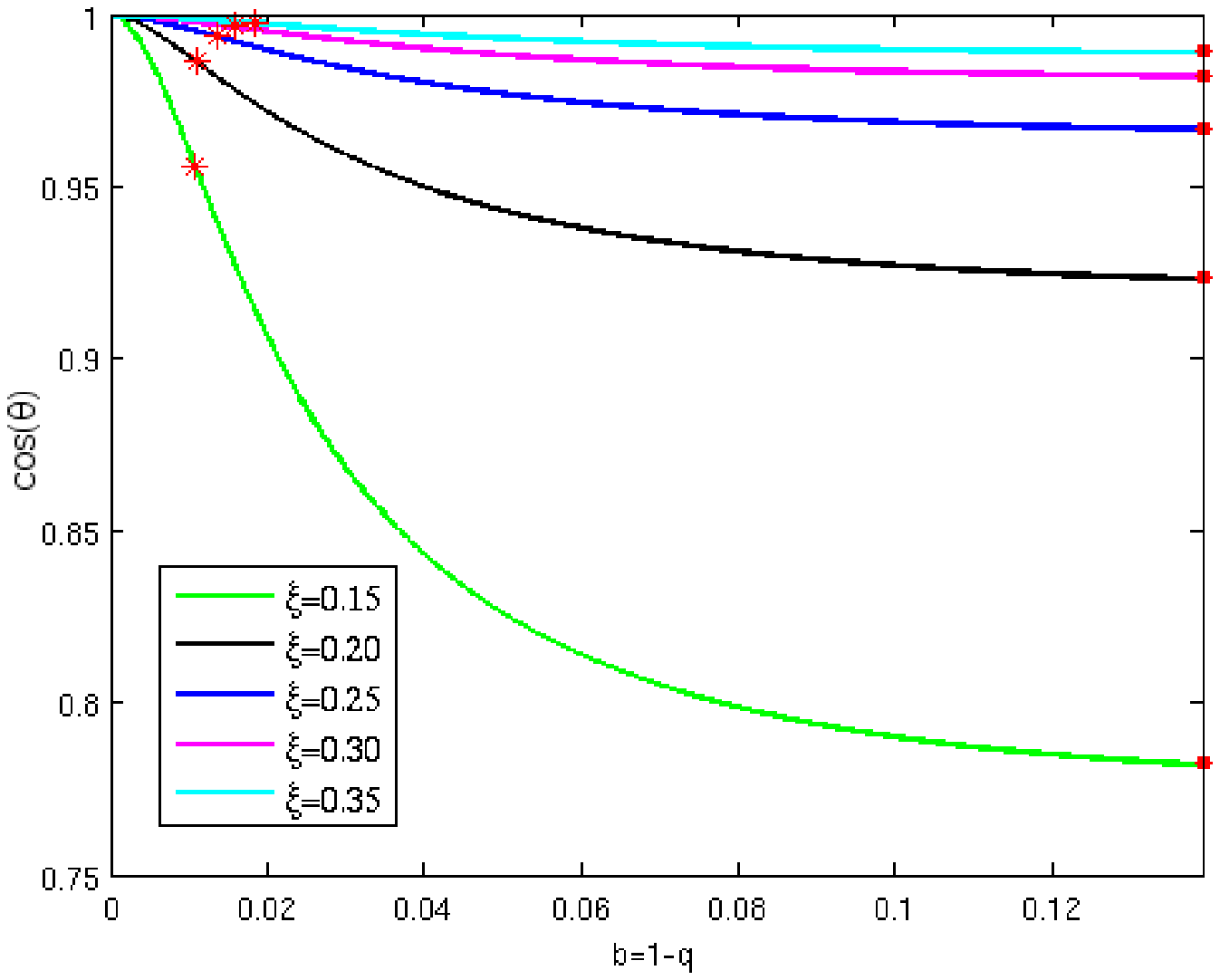}
\caption{\emph{Dependence of $\cos(\theta)$ on the error factor $b$, for the error-onto-all, discrete updates model. The size $n=20$ and $\lambda=4$ have been fixed. Each curve illustrates a different background covariance $\xi$, as shown in the legend. The inflection point on each curve was marked by a red asterisk. The inflection points are closest to zero for intermediate values of $\xi$, which agrees with the result in Figure 7. Left: ``high covariance pair" input distribution. Right: ``uniform covariance" inputs.}}
\end{figure}

Figure 5 illustrates the dependence of performance on error at various ``background correlation'' $\xi$  values in a network with 20 inputs, for the two above cases (left plot -- higher covariance pair; right plot -- high variance neuron with uniform background correlation). Since the slopes along the curves corresponding to different $\xi$ values are not simply scalings of each other, it follows that the way the performance degrades with error depends on the background correlation. For intermediate background correlation $\xi$, the output shows the highest sensitivity at very small error, while for very weak and for very strong background correlation, the maximum sensitivity appears at larger values of the error. This results in the inflection point first moving to the left as $\xi$ increases, then moving back to the right (see Figures 5a and 5b; the rightward movement is only visible at lower $\xi$ values than those shown in Figure 5b).

Figure 6 shows the dependence of performance on error for various network sizes, using fixed $\lambda$ and $\xi$ values, in both types of background correlation model. In this case the initial effect of error is very strong at large network sizes (because of synaptic crowding), but performance then reaches rather constant levels which are fairly close to the error-free level, because high background correlations tend to equalize all weights even in the absence of error. \\

We now analyze these numerical results.

\subsubsection{Model 1 - high covariance on one pair}

The principal component of ${\bf EC}$ is the unit vector pointing in the direction of $(s,s,1,...,1)^T$.

\noindent Here, the output's  ``selectivity" $s=s(n,\epsilon,\lambda,\xi)$ is given by:

$$\frac{1}{s}=1+\frac{(1-n \epsilon)(\lambda-\xi)}{z^-_{\bf EC}}$$

\noindent where $z^{-}_{EC}<0$ is the smaller root of a quadratic defined in Text S3. Once again, there is competition between the sets of preferred and nonpreferred weights.

In both models, the selectivity can be used to interpret features of the output performance with various degree
of error (see Discussion). As the explicit formula for $s$ is rather
complicated, we calculated the upper bound and the lower bound ($r(\epsilon)$), which are simpler and yet still suggest some of the main features:

$$r(\epsilon)=1-\frac{(1-n\epsilon)(\lambda-\xi)}{[1-(n-2)\epsilon](\lambda-\xi)+n(\epsilon+\xi-\epsilon \xi)}\leq \frac{1}{s} \leq 1$$

\noindent where $\displaystyle{\lim_{\epsilon \to
1/n}{r(\epsilon)}=1}$ and $\displaystyle{\lim_{\epsilon \to
0}{r(\epsilon)}=1-\frac{\lambda-\xi}{\lambda+(n-1)\xi}}$.\\

This can be compared with our other measure of output performance: the
cosine of the angle $\theta=\theta(n,\epsilon,\lambda,\xi)$ between
the principal eigenvector $(s,s,1...1)$ of ${\bf EC}$ and the
principal component of the input $(s_0,s_0,1...1)$ (where
$s_0=s(n,0,\lambda,\xi)$ is the selectivity in the absence of
error). These and other measures of performance are compared in the discussion.

$$\cos(\theta)= \frac{2ss_0+n-2}{\sqrt{2s^2+n-2}\sqrt{2s_0^2+n-2}}$$

\begin{figure}[h!]
\includegraphics[scale=0.5]{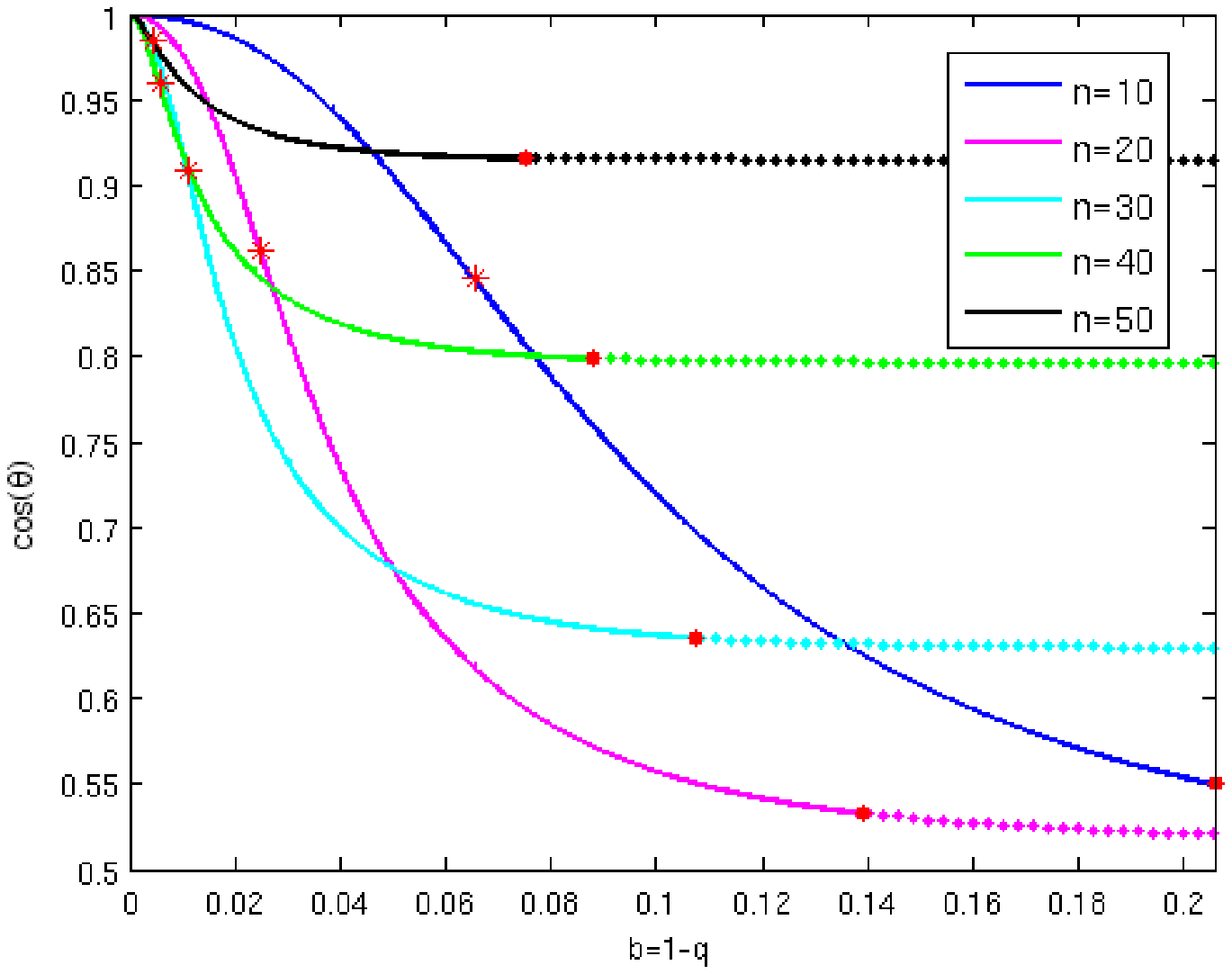}
\includegraphics[scale=0.5]{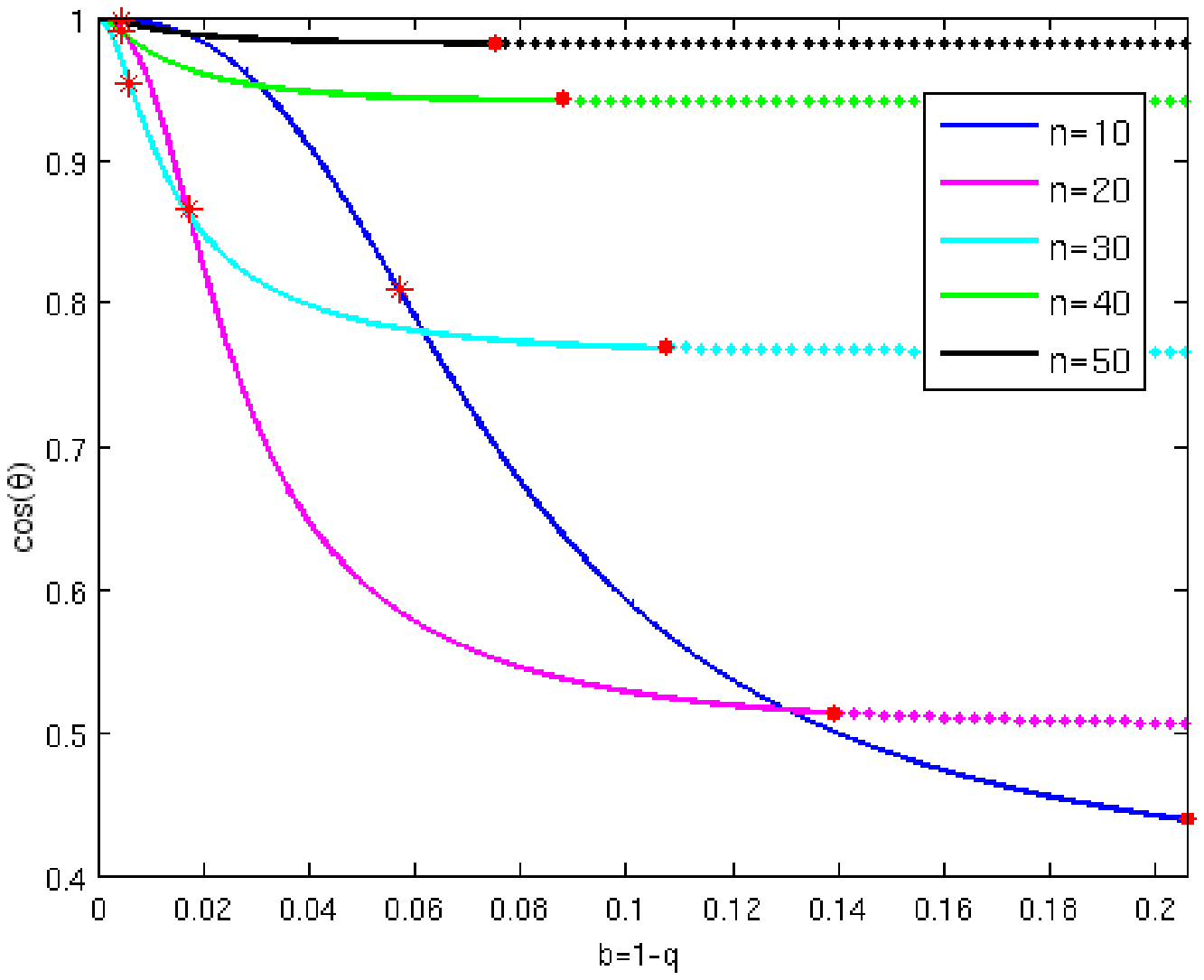}
\caption{\emph{Dependence of $\cos(\theta)$ on the error factor $b$, for the error-onto-all, discrete update model. The variances $\lambda$ and $\xi$ have been fixed to $\lambda=4$ and $\xi=0.1$. Each curve corresponds to a different network size $n$, and the inflection point on each curve is marked by a red asterisk. The infection points approach zero as $n$ gets arbitrarily large. Left: ``high-variance pair" input distribution. Right: ``uniform variance" inputs. The red dots show the trivial error values, and the curves are shown dotted beyond this point because this is a nonbiological range.}}
\end{figure}

\begin{figure}[h!]
\begin{center}
\includegraphics[scale=0.5]{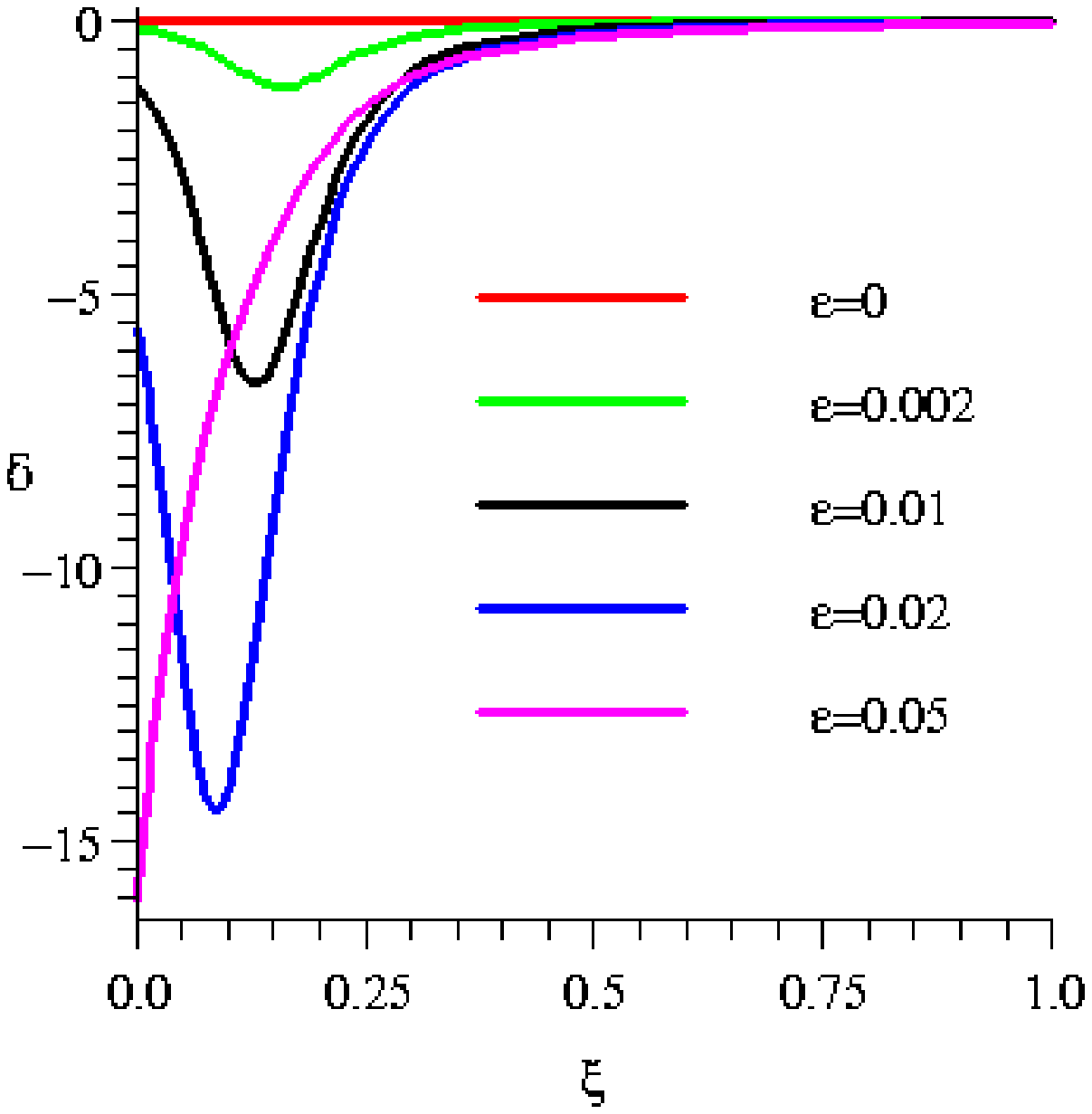}
\includegraphics[scale=0.5]{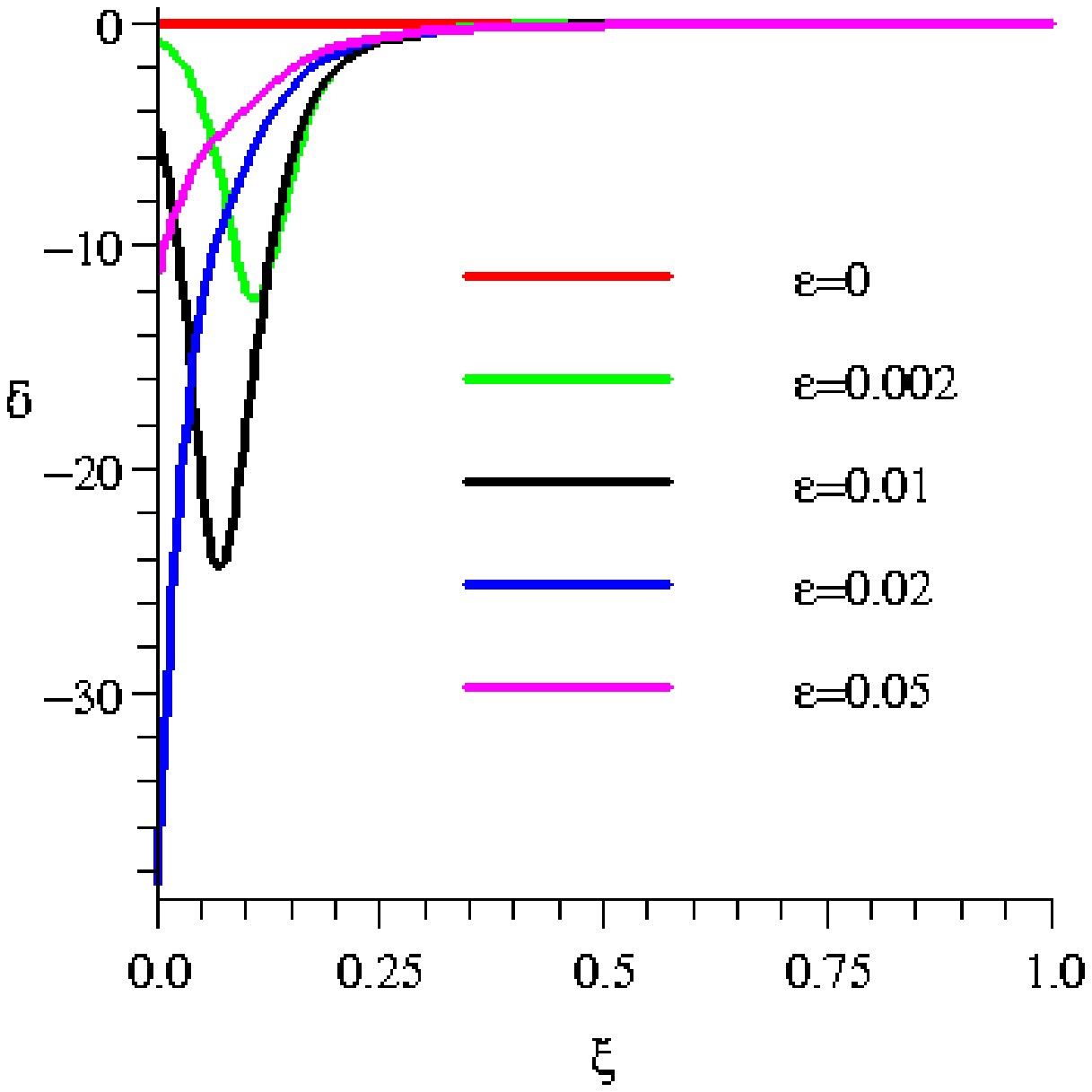}
\end{center}
\caption{\emph{Dependence of the output sensitivity $\displaystyle{\delta=\frac{\partial}{\partial \epsilon}(\cos(\theta))}$ on the covariance $\xi$ for $n=20$, $\lambda=4$, and five error values $\epsilon=0$, $\epsilon=0.002$, $\epsilon=0.01$, $\epsilon=0.02$, $\epsilon=0.04$, $\epsilon=0.05$. Each curve corresponds to a different error, as shown in the legend. The output shows the most sensitivity to $\epsilon$ at intermediate error values ($0.02<\epsilon<0.05$) and at low covariance values ($0<\xi<0.25$). Maple software was used to compute $\delta$ and generate the pictures.}}
\end{figure}

\subsubsection{Model 2 - uniform pairwise covariance}

As before, we compute the eigenvector ${\bf w}$ of ${\bf EC}$
corresponding to $\mu_{\bf EC}$. As expected, we get that ${\bf w}$ is in the direction of $(s,1,...,1)^T$.

\noindent Here, the output's  selectivity $s$ is given by:

$$\frac{1}{s}=1+\frac{(1-n \epsilon)(\lambda-1)}{z^{-}_{\bf EC}}$$

\noindent and has upper and lower bounds

$$r(\epsilon)=1-\frac{(1-n\epsilon)(\lambda-1)}{(n-1)[\xi-\epsilon(\xi-1)]}\leq \frac{1}{s} \leq 1$$

\noindent Here also $\displaystyle{\lim_{\epsilon \to
1/n}{r(\epsilon)}=1}$ and $\displaystyle{\lim_{\epsilon \to
0}{r(\epsilon)}=1-\frac{\lambda-1}{(n-1)\xi}}$.\\

The relation with $\cos(\theta)$ is given by

$$\cos(\theta)=\frac{ss_0+n-1}{\sqrt{s^2+n-1}\sqrt{s_0^2+n-1}}$$

\noindent where $s_0$ is again the selectivity for zero error.

Thus in both models $\cos(\theta)$ has a similar dependence on $b$ and $n$ (see Figure 6).

\subsection{Error sensitivity}

We define a quantity $\delta$ as the error sensitivity of the performance:

$$\delta=\frac{\partial \cos(\theta)}{\partial \epsilon}$$

Figure 7 shows Maple plots of the dependence of $\delta$ on background covariance, measured at different error rates, for the two correlated cases. $\delta$ is always negative (error degrades performance, as in the uncorrelated case), except of course for $\xi=1$, where the error-free and the erroneous equilibrium eigenvectors already have the same form. Also, $\delta$ is very small near zero error, again as in the uncorrelated case. At low error rates, adding background correlation increases the error sensitivity $\delta$. The maximum error sensitivity is greatest at intermediate error rates.

These effects reflect two opposing processes. Background correlation increases the rate of growth of all connections; from a connection's point of view it looks as though the pressure driving selective growth of one (Figure 6b) or two (Figure 6a) connections has been reduced (e.g. is equivalent to a reduction in $\lambda$ in Figure 6b). But increases in background correlation tend to make the weights more equal, synergistic with increases in error. The second effect dominates at high error values.

We looked at Maple calculations of the sensitivity of performance to changes in background covariance, $\displaystyle{\frac{\partial}{\partial \xi}(\cos(\theta))}$, at various values of $\epsilon$ and $\xi$. These can be used to understand the dependence of the sizes of the fluctuations visible in Figure 2 on parameters. We interpret these fluctuations as small deviations of the input statistics from their average values (i.e. small spontaneous transient perturbations of parameters such as $\xi$). Their amplitudes should therefore follow $\displaystyle{\frac{\partial}{\partial \xi}(\cos(\theta))}$. We found that $\displaystyle{\frac{\partial}{\partial \xi}(\cos(\theta))}$ increased as error increased, as seen in Figures 2b and 2c.

In Figure 2b independent and equal variance ``sources'' were linearly mixed to generate correlated random vectors used as inputs to the erroneous Oja rule. These correlations act as a ``background'' which tends to equalize the weights even in the absence of error, so adding error has relatively little effect.

\subsection{Other models and extensions}

Here we consider an input distribution such that the variance is higher, but uneven, on two of the components, while the covariance is uniform (and possibly zero). The correlation matrix will be of the form:

\begin{eqnarray}
{\bf C}=\left( \begin{array}{ccccc}
       \lambda_1&\xi&\xi&\cdot&\xi \\
       \xi&\lambda_2&\xi&\cdot&\xi \\
       \xi&\xi&1&\cdot&\cdot\\
       \cdot&\cdot&\cdot&\cdot&\xi \\
       \xi&\xi&\cdot&\xi&1
       \end{array} \right)
\end{eqnarray}

\noindent with $\lambda_1>\lambda_2>1>\xi$.

The modified correlation matrix ${\bf EC}$ has the eigenvalue $(Q-\epsilon)(1-\xi)$, with multiplicity $n-3$. The other three eigenvalues $\mu_1$, $\mu_2$ and $\mu_3$ are distinct and lie respectively within the intervals: \\

$(Q-\epsilon)(1-\xi)<\mu_1<(Q-\epsilon)(\lambda_2-\xi)$

$(Q-\epsilon)(\lambda_2-\xi)<\mu_2<(Q-\epsilon)(\lambda_1-\xi)$

$\max \{(Q-\epsilon)(\lambda_1-\xi), n \xi+(1-\xi)+\epsilon(\lambda_1+\lambda_2-2)\}<\mu_3< n(\xi+\epsilon-\epsilon \xi)+\epsilon(\lambda_1+\lambda_2-2)$\\

\noindent Clearly, $\mu=\mu_3$ is always the unique maximal eigenvalue of ${\bf EC}$.

In the case of uncorrelated inputs, for example, $\epsilon=0$ corresponds to $s_1=1$ and $s_2=0$ (the maximal eigenvalue $\mu=\lambda_1$, and its corresponding eigenvector is the first element $(1, 0, ... 0)$ of the standard orthonormal basis in $\mathbb{R}^n$. As the error increases from $\epsilon=0$ to $\epsilon=1/n$, the eigenvector $(s_1, s_2, 0....0)^T$ (see Figure 8) evolves such that the ratio $s_1/s_2$ decays very dramatically from $\infty$ (when $\epsilon=0$) to finite values. When $\epsilon \to 1/n$ (the trivial value) all weights equalize and thus $s_1/s_2 \to 1$ (Figure 8b). Thus in a situation where 2 highly (but inevitably unequally) active inputs are to be selectively wired by Hebbian learning, the presence of error actually promotes the desired outcome, at least in the large $n$ case.

\begin{figure}[h!]
\includegraphics[scale=0.5]{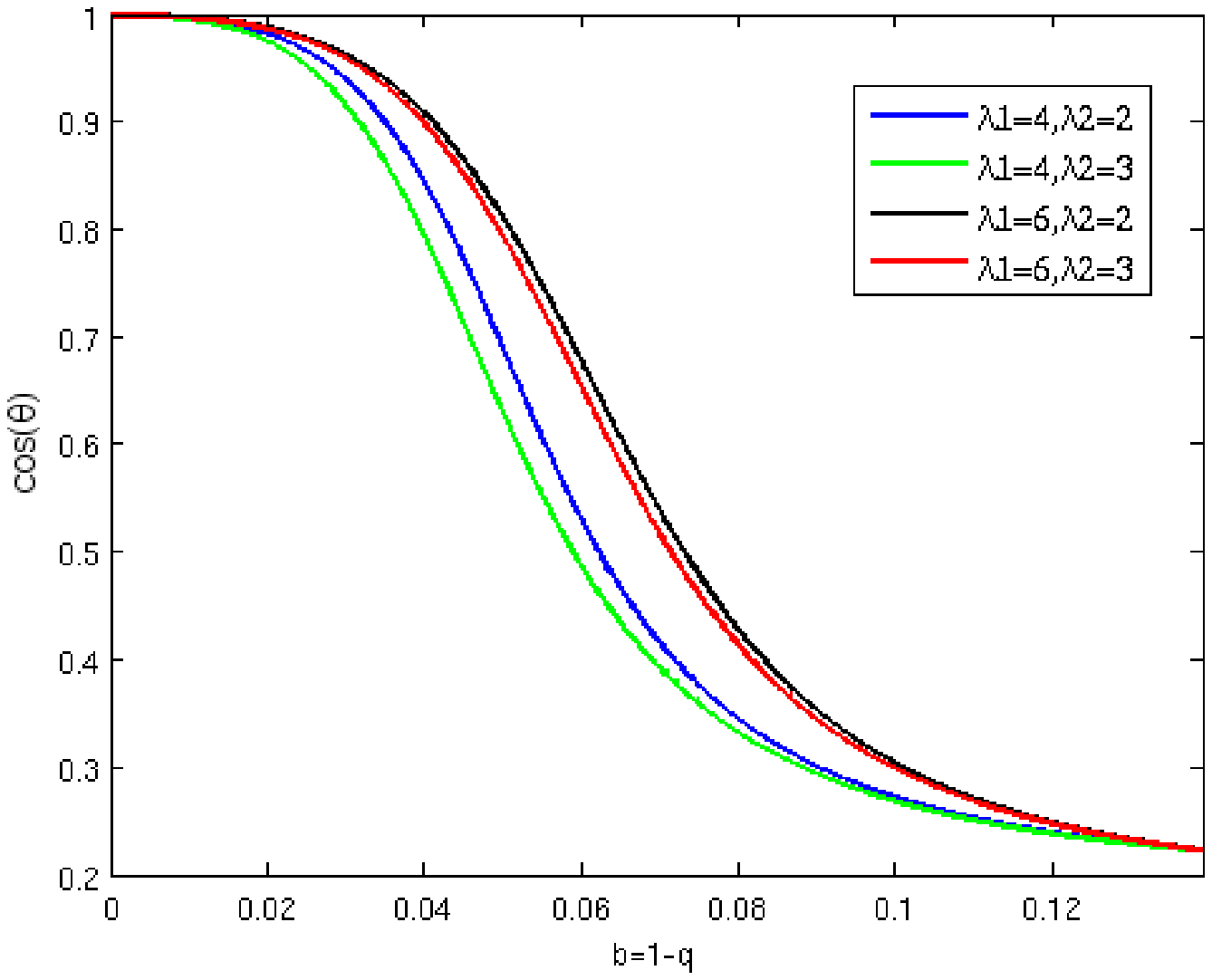}
\includegraphics[scale=0.5]{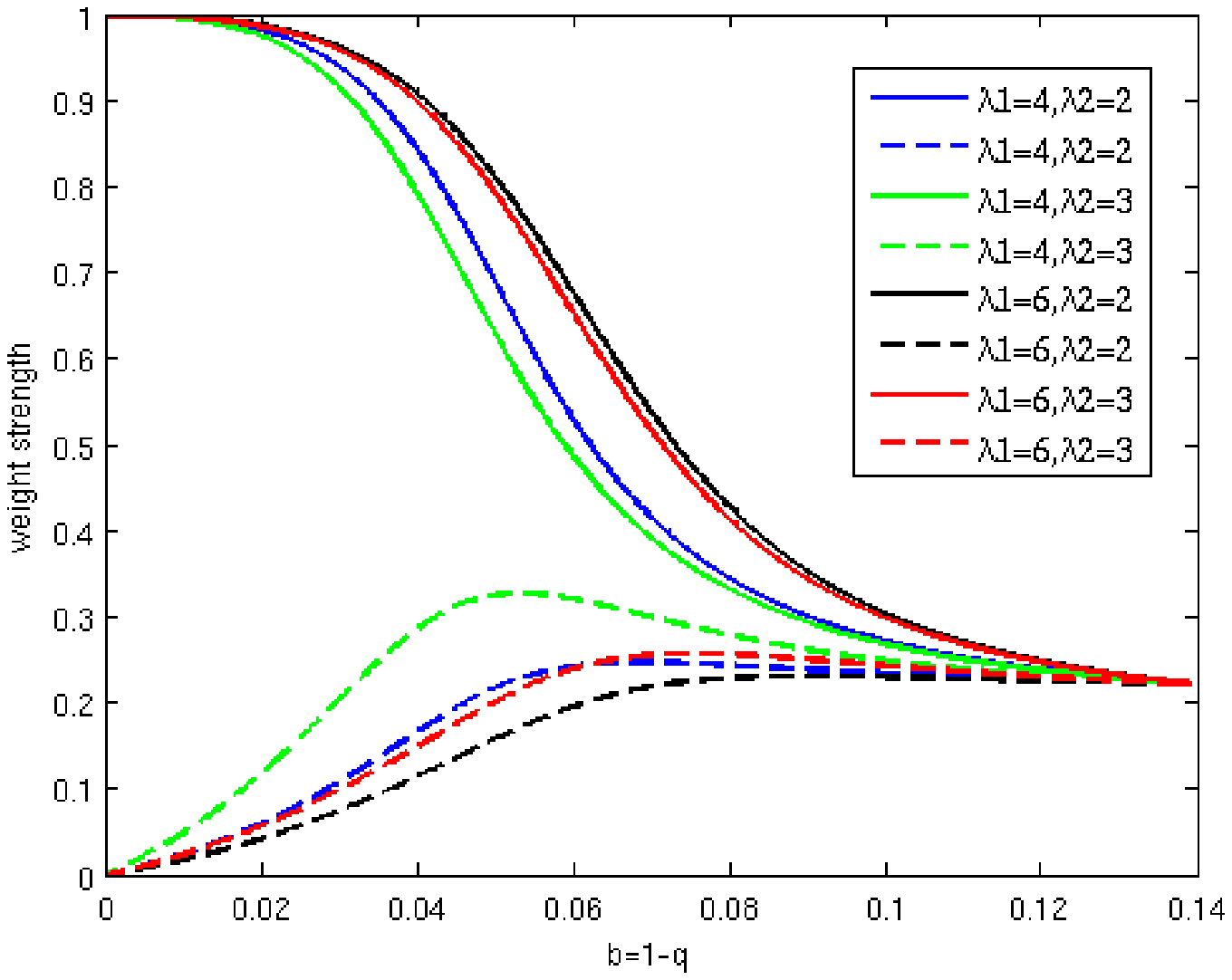}
\caption{\emph{Error-onto-all, discrete update model for $n=20$ cells receiving uncorrelated inputs with $\lambda_1>\lambda_2>1$. Left: Dependence of $\cos(\theta)$ on the synaptic error $b$, shown as $b$ increases from zero to the trivial value $b_0(n)=1-1/\sqrt[n]{n}=1-1/\sqrt[20]{20} \sim 0.14$. The trivial error $b_0(n)$ equalizes all weights and makes $\cos(\theta)=1/\sqrt{n}=1/\sqrt{20} \sim 0.22$, independently of the distribution variances $\lambda_1$ and $\lambda_2$. Right: Evolution of the normalized weights $\frac{s_1}{\sqrt{s_1^2+s_2^2+(n-2)}}$ and $\frac{s_2}{\sqrt{s_1^2+s_2^2+(n-2)}}$ with respect to $b$. As $b$ increases from zero to $b_0(n)$, the weights equalize and the ratio $s_1/s_2$ drops from $\infty$ to $1$.}}
\end{figure}

When the inputs are correlated, the dependence of $\mu$ on parameters is more complicated. The eigenspace of $\mu$ is the direction $(s_1, s_2, 1,...1)^T$, where the selectivities $s_1$ and $s_2$ themselves depend, via the eigenvalue $\mu$, on all the system parameters:

$$\frac{s_1}{s_2}=1+\frac{(Q-\epsilon)(\lambda_1-\lambda_2)}{\mu-(Q-\epsilon)(\lambda_1-\xi)}$$

\noindent It is easy to observe that, as $\epsilon \to 1/n$ (the trivial error value), $Q-\epsilon \to 0$, hence $s_1/s_2 \to 1$. As in the uncorrelated case, weights tend to equalize as the error gets close to the trivial value (see Figure 9). However, the slope of the decay of $s_1/s_2$ is different from the uncorrelated case, since $s_1/s_2$ is always finite when the inputs are correlated with $\xi>0$, even for zero $\epsilon$.

\begin{figure}[h!]
\includegraphics[scale=0.5]{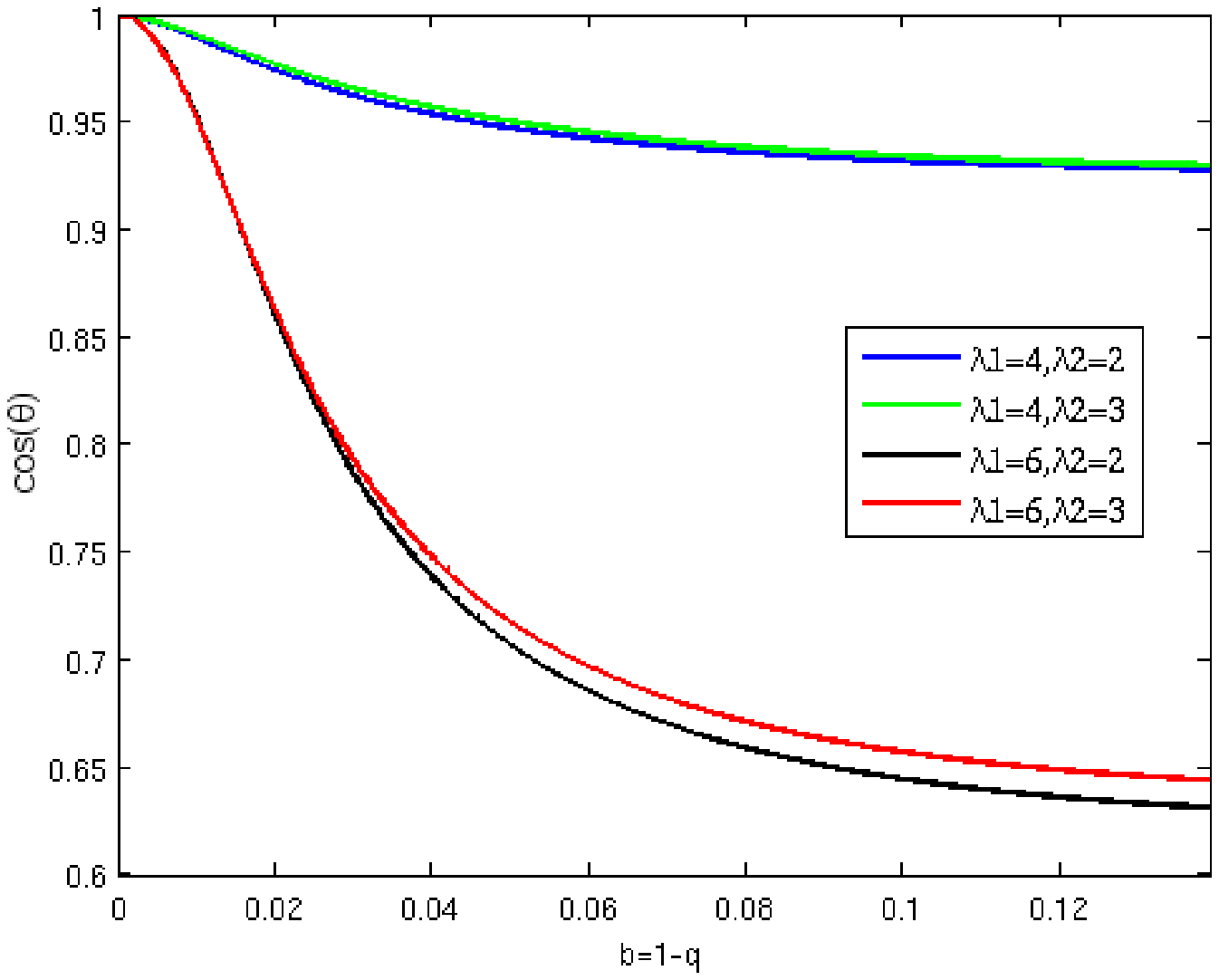}
\includegraphics[scale=0.5]{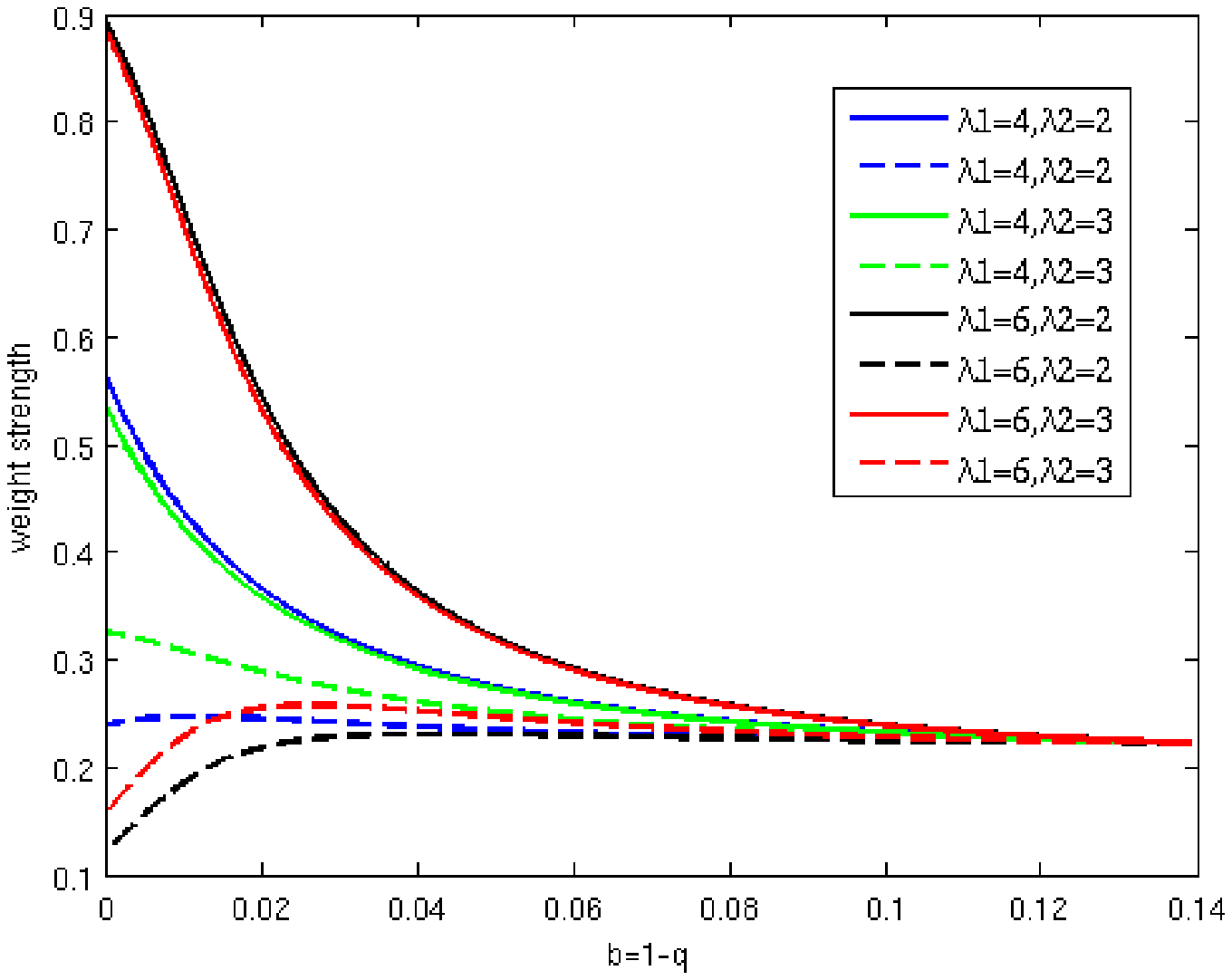}
\caption{\emph{Error-onto-all, discrete update model for $n=20$ cells receiving correlated inputs with variances $\lambda_1>\lambda_2>1$ and small uniform covariances $\xi=0.2$. Left: Dependence of $\cos(\theta)$ on the synaptic error $b$, shown as $b$ increases from zero to the trivial value $b_0(n)=1-1/\sqrt[n]{n}=1-1/\sqrt[20]{20} \sim 0.14$. The trivial error $b_0(n)$ equalizes all weights, but $\cos(\theta)$ varies at $b_0(n)$. Since the principal component of ${\bf C}$ varies with parameters, so will the angle $\theta$ at the trivial error value. Right: Evolution of the normalized weights $\frac{s_1}{\sqrt{s_1^2+s_2^2+(n-2)}}$ and $\frac{s_2}{\sqrt{s_1^2+s_2^2+(n-2)}}$ with respect to $b$. As $b$ increases from zero to $b_0(n)$, the weights equalize and the ratio $s_1/s_2$ drops from an initial finite, parameter-dependent value to $1$.}}
\end{figure}

\begin{figure}[h!]
\begin{center}
\includegraphics[scale=0.55]{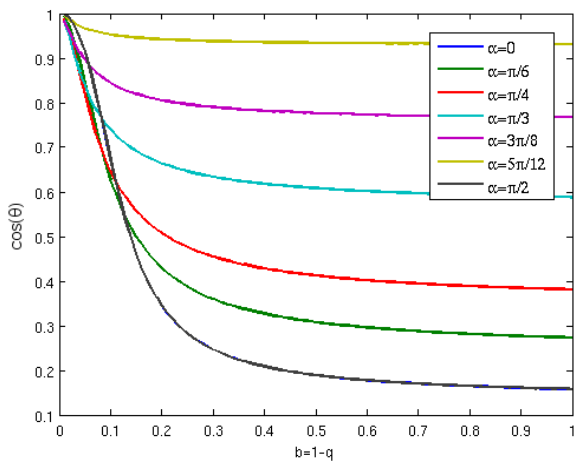}
\caption{\emph{Oja network learning a distribution of correlated inputs obtained by rotations of $n$-dimensional normally-distributed vectors.Here $\lambda=10$ and $n=40$. The amount of rotation alpha was varied as shows in the inset.}}
\end{center}
\end{figure}

Although these results are not general, they seem to apply to various other situations with increasing degree of background correlation (e.g. Figure 2). Similar behavior can be observed, for instance, in an Oja network learning from correlated inputs obtained by rotations of $n$-dimensional normally-distributed vectors. Once again one sees that as correlations increase the inflexion points in the performance-error plots shift to the left and then to the right (Figure 10; compare to Figure 5 and Figure 2b), confirming that at low error introducing correlation increases error sensitivity.

\section{Discussion}

\subsection{Overview}

The present study forms part of our ongoing effort ~\cite{Adams1}~\cite{Adams2} to evaluate a novel, sweeping but ultimately prosaic hypothesis about the neural basis of ``mind''. If sophisticated brains are machines for learning about the structure of the world~\cite{Barlow}~\cite{Marr}~\cite{Dayan}, then we propose that a key issue becomes the accuracy of synaptic learning, just as a key issue underlying Darwinian evolution (a form of ``molecular intelligence''~\cite{Adami}) is replication accuracy~\cite{Tarazona}~\cite{Swetina}~\cite{Eigen1}. In both cases physical limits to biological accuracy~\cite{Bialek} would set the amount of compressed information that can be stored. In this view, the neocortical microcircuit would be a device allowing highly accurate synapse adjustment, and thus learning from weak higher-order correlations~\cite{Adams2}. Thus, just as the core machinery of all cells is devoted to accurate replication~\cite{Watson}, even though different types of cells perform additional complex functions, the core neocortical circuitry would allow accurate learning, and the remaining more variable circuitry which is traditionally studied would be merely a specialized, though interesting, add-on.

 Our results here provide background for the exploration of this simple and apparently powerful idea, but before reviewing them in detail, we briefly consider why the idea has been completely overlooked. There seem to be several interacting factors (apart from the inadequacy of our previous explanatory attempts).

 First, there are widespread assertions that Hebbian adjustments are completely accurate~\cite{Sabatini1}~\cite{Koch} because the underlying calcium signal does not spread beyond the spine head~\cite{Sabatini2}. Such assertions, assisted by misleading colorizations of ambiguous data~\cite{Purves}, run counter to the necessity that the spine neck resistance (electrical or diffusional) must be relatively low at powerful synapses, as well as to experimental evidence~\cite{Tao}~\cite{Engert1}~\cite{Harvey}. Second, almost all the relevant neural network theory~\cite{Dayan}~\cite{Haykin}~\cite{Herz} has been developed using weight-specific learning rules, partly for analytical simplicity, partly because of the widespread currency of the synapse isolation hypothesis, partly because most technical work naturally focuses on proving that algorithms work, and partly because physical errors do not occur in serial computer implementations of such algorithms. It is likely however that the analog, massively parallel implementations that may be required to scale neural net algorithms to real-world problems~\cite{Likharev} will also suffer from the learning accuracy problem.

Second, biology can usually safely be sloppy, but there is one glaring exception: DNA replication is almost miraculously accurate. But this exception is the only relevant parallel, since Darwinian evolution and neural learning are both adaptive processes which store information about the environment based on repeated interactions~\cite{Baum}~\cite{Volkenstein}~\cite{Adami}.

Third, the view that most neocortical circuitry may not be devoted to specialized information-processing tasks, while not intrinsically absurd, is difficult to swallow by those who focus on elucidating such information-processing.

The current paper extends our initial attempts~\cite{Cox}~\cite{Adams1} to exploring the effects of Hebbian inaccuracy in neural network learning. We selected unsupervised learning for two reasons. First, it is likely that the vast majority of learning is unsupervised, simply because labeled examples are relatively rare. Second, many studies of supervised learning use ``real-world'' data that does not conform to a simply-describable statistical model (of course, that is the whole point of doing supervised learning).  In the present paper we explore the classic Oja model of a neuron as a Principal Component Analyzer~\cite{Oja1}~\cite{Diamantaras}. In the next section we justify this choice.

\subsection{Is the Oja Model Biologically Realistic?}

Since our focus is on biological realism, it may seem odd to study a model which is apparently almost as formal and unbiological as one can imagine. However, our goal is not to construct a biologically-detailed model of learning by realistic neurons, but to better understand one specific aspect of realism that has hitherto been neglected: the possible inspecificity of the learning rule. Making other aspects more realistic would unnecessarily complicate the task (and might prevent analytical treatment). In this section we argue that the Oja rule is not as biophysically unrealistic as first appears. However, we do not address the issue of whether brains actually do PCA; it seems likely that at least in the visual system other more locally decorrelating representations are developed~\cite{Laughlin}~\cite{Atick1}~\cite{Atick2}~\cite{Dayan}~\cite{Bell}, insofar as these are learned, crosstalk would probably have similar effects.

The first obvious simplification in the Oja rule is that both pattern elements and weights are allowed to be negative. However, if only positive patterns are allowed, the Hebbian part of the rule would always be positive, and so the rule only requires LTP for this part. Conversely, the normalizing part of the rule would always be negative, and would only require LTD. Biological mechanisms for LTP and LTP are considered below. Furthermore, it seems that in the brain the negative and positive parts of signals are represented using different neurons (e.g. on and off cells in the retina and thalamus); this means that even though the two halves of the Oja rule would operate biologically with fixed and opposite polarities (LTP and LTD), the overall effect of the biological implementation would be the same as the original Oja rule, which allows either polarity in both parts of the rule.

The next simplification in the Oja rule is that the temporal relations of incoming patterns are ignored. While time is often of great biological importance (music and movies) sometimes it is less so (painting and sculpture), and the extent to which brain systems specialize for spatial or temporal resolution varies. Here the crucial point is that it is likely that adding time-dependence to a learning rule (i.e. STDP) would be unlikely to make it easier to implement at the physical, synapse, level. The basic problem is that if plasticity has to be triggered by calcium over a narrow time window, then the calcium signal itself must be larger (given that most biological binding processes are diffusion controlled) which will place even greater demands on pumps and buffers. Viewed from a different angle, increased speed can only be achieved by decreased affinity (and hence worsened selectivity) in the diffusion-controlled regime. There is also the difficulty that the mathematics of time-dependent learning rules are more complicated, especially if the spread of errors between synapses has to be represented by a time-dependent ``error matrix''. In the simplest case, multiplications would be replaced by convolutions.

Closely related to this is the implicit ``rate-coding'' in the Oja model. However, this seems justified if temporal order is unimportant; the mean rate can be thought of as a firing probability over some suitable small time window  ($\sim$ the membrane time constant), and the behavior of a synapse in response to a repeated interleaved stochastic spike pattern should, as a ``meanfield'' approximation, be identical to the deterministic response to continuous-valued inputs. This leads to the implementation of the learning rule as coincidence-detection. The standard (though rarely-articulated) view is that the Hebbian multiplication is biologically implemented by making tiny increases in strength whenever pre- and post-synaptic spikes  fall within a small time window (presumably comparable to the membrane time constant). This produces ``stochastic multiplication''~\cite{Likharev} of firing probabilities, if the input and output fire independently (as implied by the underlying assumption that firing occurs in a Poisson fashion). Whether ``tiny'' reflects small changes produced by every coincidence or bigger changes generated stochastically by only some coincidences is discussed below).

It seems to us that within the constraints of the atemporal viewpoint, the match between the Hebbian multiplication in the Oja rule and the actual machinery of coincidence-detection is pretty good. There is direct experimental evidence that a back-propagating spike that arrives near the peak of the activation of spine-head NMDARs does lead to a more-or-less synapse-specific extra calcium-influx~\cite{Nevian1} and also to an increased probability of strengthening~\cite{Kalisma}~\cite{Markram1}. It is widely thought that this reflects Mg-unblocking of the NMDAR~\cite{Letzkus} though no adequate quantitative model has been constructed, because of uncertainties about the calcium current -- voltage relationship~\cite{Wollmuth}. Furthermore, this extra influx will be smaller if the timing of the back-propagating action potential is suboptimal, and since the time course of NMDAR activation is similar to the time-course of unitary epsps~\cite{Koch2}, there seems to be a rough matching between the time-course of the increased calcium entry and the time course of the increased firing probability following single-axon (unitary) activation. If the unitary firing probability has a more rapid time course than the corresponding unitary epsps, this could be compensated by complications in the blocking details~\cite{Kampa}~\cite{Letzkus}.

Our model uses continuous-valued weights; in the ``discrete'' update model, outlined in the Text S2, although the updates themselves are all-or-none (in agreement with experiments~\cite{Petersen}~\cite{OConnor}~\cite{Bagal}) continuous weights are still obtained in the limit of larger numbers of synapses (or by time-averaging over smaller numbers of synapses). The approximate discrete model we mostly used in the Results becomes exact in the low error limit; conversely, the continuous model we sometimes also use (e.g. Figure 2) approaches the exact discrete model at high error or large numbers of synapses.

One difficulty, surprisingly, is the linear form of the Hebb part of the Oja rule. In the model of calcium-mediated inspecificity presented in Text S2, we introduce a parameter $h$ which reflects the ``Hill coefficient'' for the activation of CaMKinase~\cite{DeKoninck}, which measures the number of calcium ions required to activate the enzyme. Because CaMKinase comes as two different genes each of which leads to several differentially spliced variants (somewhat like the BK channel proteins that mediate the wide kinetic range of hair cell electrical turning~\cite{Gaertner}~\cite{Jones}), it is likely that this parameter, as well as Ca affinity, can be tailored at different types of synapse for different functions. Nevertheless, $h$ is likely to be always greater than one, meaning that the change in the strength of a synapse as a result of spine head calcium increases (whether generated locally or secondary to increases at other synapses) is likely to be a nonlinear function of [Ca]$_i$. Conversely, in the simplest case, the size of the calcium signals would be a linear function of pre- and postysynaptic firing. This would lead to a nonlinear Hebb rule, which would not in general be solely driven by the input covariance matrix. The representation would therefore (for approximately Gaussian input statistics) be suboptimal. (The possibility that such nonlinearities are exploited to reduce error-sensitivity is discussed below).

One way that neurons could handle this problem is to cancel out the intrinsic nonlinearity of the calcium transduction machinery with a suitable ``inverse'' nonlinearity in the calcium induction machinery, so that overall the strength change is roughly linearly related to the product of the firing rates. However, strict cancellation would probably be difficult to achieve. What is needed is a demonstration that an approximately linear Hebb rule learns approximately the first PC. In the general case, this may be elusive, because the outcome of nonlinear learning depends on the details of the input statistics (unlike the linear case, which is only sensitive to the pairwise statistics). In the simplest case, where the input statistics are exactly Gaussian, nonlinear learning behaves linearly since all the higher order statistics cancel out~\cite{Hyvarinen}. So to the extent that the inputs that the brain encounters are Gaussian (probably a good approximation at the earliest stages of sensory processing), strict cancellation might not be needed. Furthermore, adding learning inspecificity has the effect of linearising the learning rule (indeed, is can completely prevent learning of high order correlations (Cox and Adams, unpublished). In conclusion, though there are important nonlinearities in the mechanisms that translate spikes to strength changes, the overall rule can be linear in its effect, through a combination of low $h$, cancellation, Gaussian statistics and inspecificity. The problem is likely not to be in early linear learning, but in later nonlinear learning.

A final issue here is whether the learning should be done after every pattern (``on-line'') or after many patterns have been accumulated (``batch-model''). In the present work we used the on-line recipe, but presumably a batch mode would work almost as well, provided the number of patterns that are ``batched'' are small compared to the total number of patterns. However, if biological weight adjustments are stochastic and all-or-none~\cite{Petersen} the distinction may not be important. If the probability of a weight change following a spike coincidence is very small (as it must be if learning is slow), it follows that the only way to reliably produce a weight change as a result of a series of coincidences is if the small probabilities accumulate. Thus a synapse should contain a ``register'' of past coincidences; each coincidence would increment the register by one step, and when the counts in the register exceed a threshold the weight would be increased. (There could also be a stochastic version where each coincidence increments the register by a variable amount that is equal to one on average). In such models the register contents should also slowly decay (or else be decremented by anticoincidence); the learning rate would be set by both the threshold and by the decay rate.

Experimental studies of LTP at single synapses strongly suggest the batch model, since many repeated ``pairings'' of correctly timed pre- and postsynaptic spikes (or in some cases, depolarisations) are required to reliably induce LTP, which occurs in an all-or-none manner~\cite{Markram2}~\cite{Petersen}. However, averaged over the many synapses comprising a connection, the overall outcome would be the multiplicative hebbian rule. A simple mechanism for such batching would be if the coincidence-induced calcium increase at a synapse activated (by binding of calmodulin) some fraction of its CaMKinase molecules; after each calcium pulse, Ca-Calmodulin would dissociate but leave some of the kinase molecules phosphorylated; with successive pulses  eventually enough would be activated that the entire set of CaMKinases would autophosphorylate, triggering strengthening~\cite{Lisman}~\cite{Lisman2}~\cite{DeKoninck}.

The biological implementation of the normalizing (LTD) part of the Oja rule is less clear. This part of the rule is elegant, since the basic normalization step (division by the Euclidean norm of the weight vector) leads, in the second-order approximation, to a purely local, online, rule. However, there are two nontrivial biophysical requirements: (1) the calculation of  $y^2$ (2) the multiplication by $\mbox{\boldmath $\omega$}$. Recent work in neocortex~\cite{Sjostrom1}~\cite{Sjostrom2} suggests that LTD occurs in the following way: backpropagating spikes lead to a synapse-related calcium signal that triggers endocannabinoid release from the local dendrite (perhaps from the spine itself); the endocannabinoid then diffuses back to the presynaptic specialization, where it activates a G-protein-coupled endocannabonoid receptor; if there is near simultaneous activation of presynaptic NMDARs by spike-release glutamate, transmitter release is depressed. It now seems likely that previously favored models, where the level of the spine calcium achieved by LTP -- or LTD -- inducing stimuli produces determines the sign of the strength change~\cite{Lisman2}~\cite{Shouval} is wrong~\cite{Nevian2}.

This new picture of LTD seems well suited to meet the two biophysical requirements of the normalizing part of the Oja rule (and in this sense the rule would be more than a formal description). The calcium-dependent endocannabinoid enzyme  triggered by calcium  entering through voltage-dependent channels activated by backpropagating spikes would implement $y^2$, and the multiplication would be achieved by the requirement for simultaneous activation of the NMDAR. The dependence on $\mbox{\boldmath $\omega$}$ could be achieved in two ways: the endocannabinoid signal might be proportional to the postsynaptic strength of the synapse, or the extent of activation of the presynaptic NMDAR could depend on the amount of glutamate released, which would depend on the extent of the active zone, which is known, in the long term, to adjust to match the psd area (and hence presumably the synaptic strength). Thus the synaptic strength would slowly adjust, by a combination of matched but distinct post- and pre-synaptic adjustments, to reflect the arriving spikes, in the way required by the Oja rule.

This background is necessary to discuss the important issue of the accuracy of the normalizing part of the Oja rule. Clearly, if LTD is triggered presynaptically by a retrograde messenger, one must consider the possibility of extracellular LTD crosstalk. If the LTD part of the rule is implemented as described above, errors in the diffusioin of retrograde messenger to different synapses on the same neuron would not matter, although diffusion to synapses located on other neurons would matter. This problem is avoided because the read-out of the weight by the requirement for presynaptic NMDAR activation by simultaneously released glutamate, is itself dependent on the occurence of appropriately-timed presynaptic spikes. If instead (and apparently unbiologically) the weight is read out postsynaptically, and the combined signal $y^2 \mbox{\boldmath $\omega$}$ is then retrogradely back-propagated to the ``correct'' presynaptic structure, diffusion of the retrograde signal would cause normalization errors. In a nutshell, this could be modeled by adding a new error matrix $F$ so the averaged rule would become

$$\Delta {\bf w} = {\bf F}({\bf F^{-1}EC w} - ({\bf w^Txx^Tw}){\bf w})$$

At first glance it appears that the normalization errors could
``cancel out'' the Hebbian errors if ${\bf F}$ is appropriately
matched to ${\bf E}$ (i.e. both "error-onto-all" with adjustment of
quality). Such cancelation would correspond to a weight erroneously
``forgetting'' exactly what it erroneously learns for each pattern.
The problem is that while the averaged values of ${\bf E}$ and ${\bf
F}$ are simple and closely related the instantaneous values ${\bf
{\cal E}}$ and ${\bf {\cal F}}$ can be, at least locally, quite
different, because one involves intracellular diffusion and the
other extracellular diffusion. Furthermore, the stability of the
algorithm will also be affected.

\subsection{Performance of PCA. Mutual Information}

In this paper we used the cosine of the angle between the learned and the correct eigenvector as a simple measure of performance. A more natural measure would be the mutual information between the input vectors and output scalar, but this requires knowledge of the distributions of the components of the input vectors. In the simplest case the distributions will be Gaussian, and it is known that in this case the PC is the optimal representation of the inputs by a single scalar (the mutual information is given by $0.5 y^2 (\log y^2)$ plus a constant; since the output variance is maximal when the weight vector  parallels the PC, so is the MI). In the erroneous case, the variance is less, and so is the MI. We found that the curves for the decline of MI as a function of parameters such as error, $n$ or variance/covariance for Gaussian inputs (Figure 11) were very similar to the curves for $\cos(\theta)$ shown here; we preferred to use $\cos(\theta)$ as a performance measure because it does not depend on input statistics.

\begin{figure}[h!]
\includegraphics[scale=0.5]{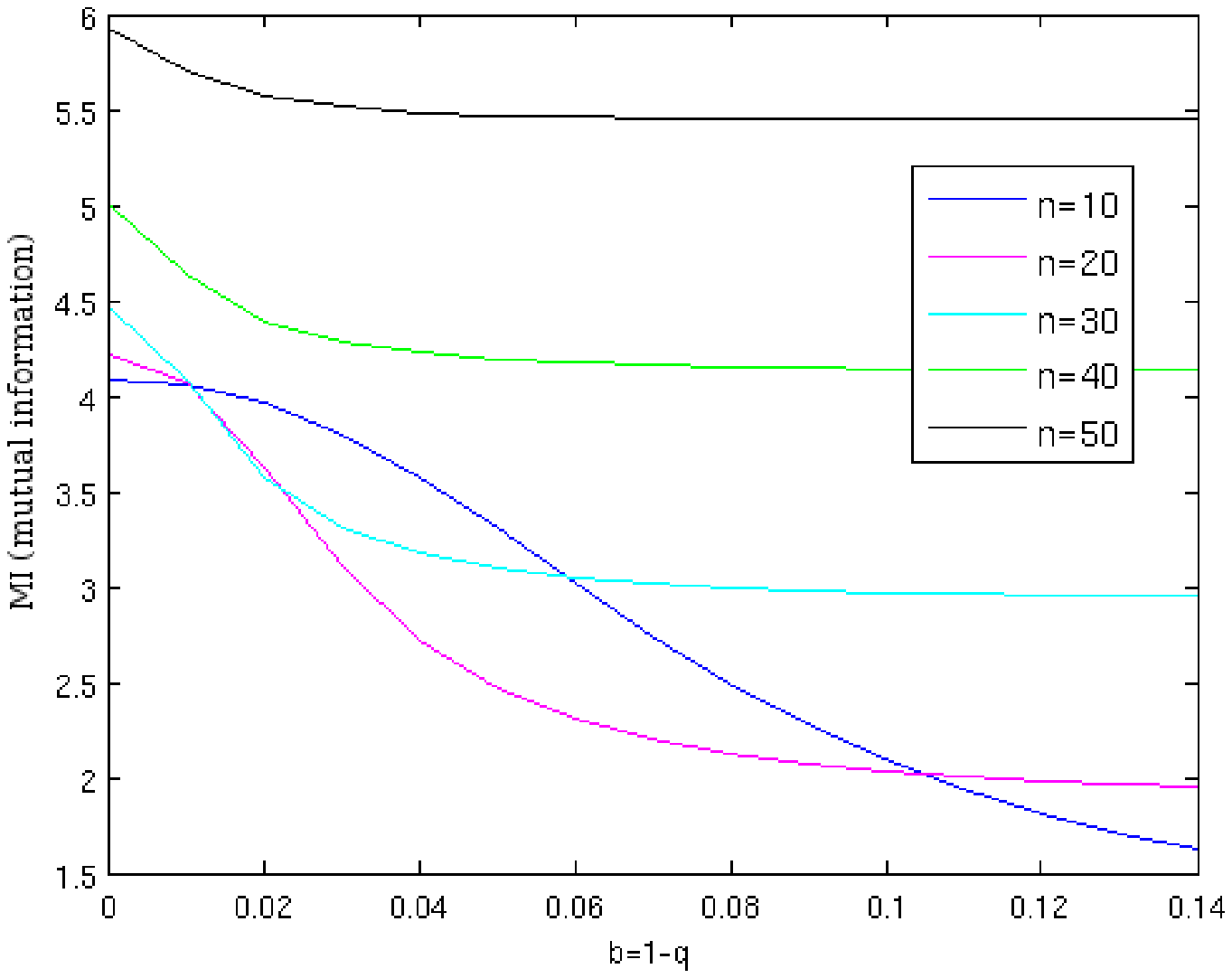} \quad \quad
\includegraphics[scale=0.5]{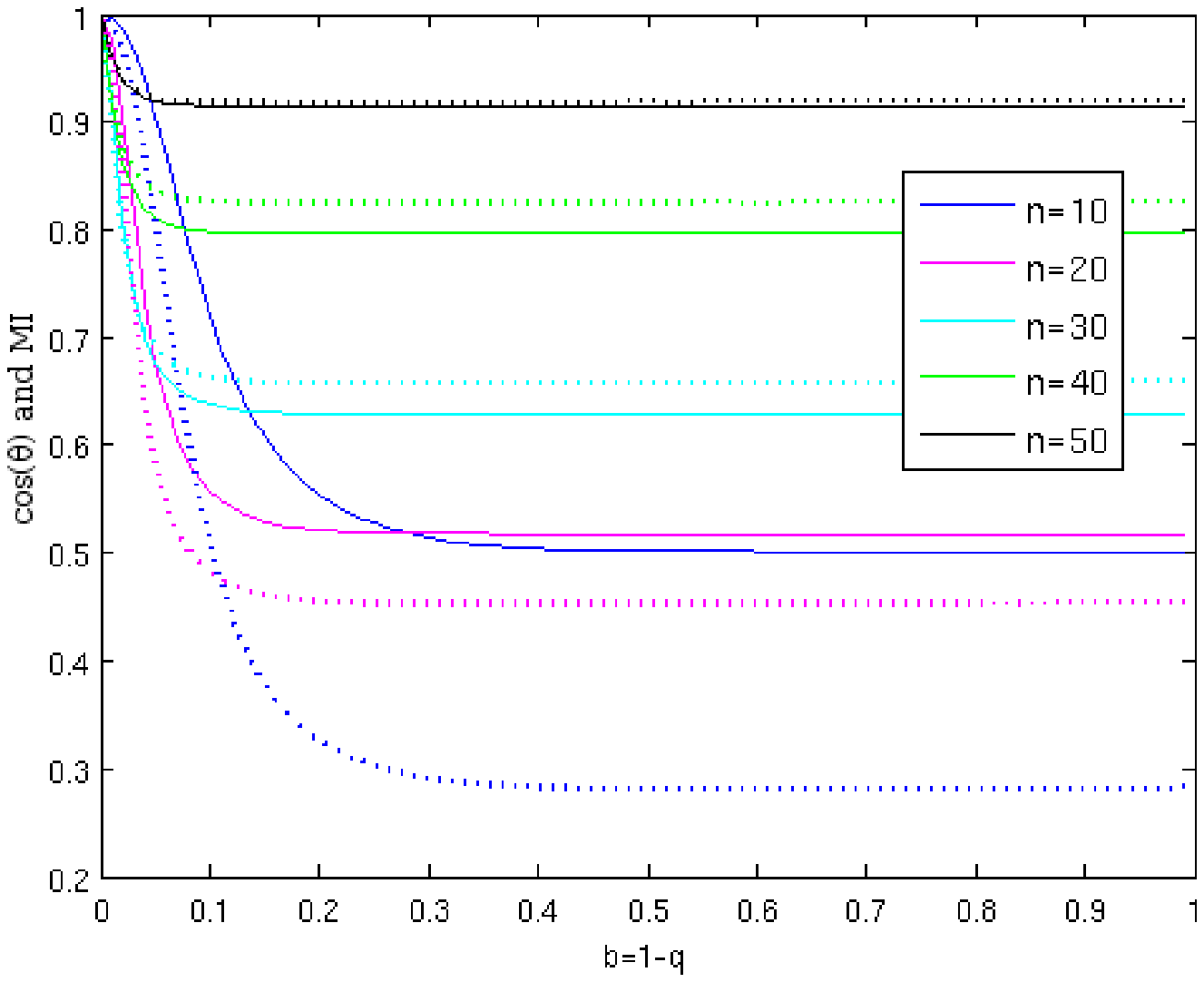}
\caption{\emph{Mutual information (MI) at different error values and network sizes, for Gaussian inputs. Left: the plots show MI as function of error for different network sizes. The MI depends on output variance; since all the inputs contribute to the output variance, the MI increases with network size. Because error increases, the relative contribution of the poorly correlated inputs to the output variance, MI decreases with error. Right: The MI (dotted lines) at different error rates is expressed as fraction of the MI at zero error, and compared with $\cos(\theta)$ at different errors (solid lines) for various network sizes. In all cases error decreases MI, and when $\cos(\theta)$ is small, it closely tracks the MI.}}
\end{figure}

\subsection{Biophysics of error}

In early experiments it was found that LTP was connection-specific if the connections were located far apart on the dendrites~\cite{Andersen}; more recent work showed that LTP induced at a localized set of synapses spreads to unstimulated synapses that are less than 50 microns away; no spread was seen beyond 70 microns~\cite{Engert1}. Uisng a somewhat different protocol, such spread was not seen even at 10 microns separation~\cite{Chevaleyre}. Even more recent work examined the possible spread of LTP induced at single synapses; no changes were detectable at even closely neighboring synapses ($\sim$ 1um~\cite{Matsuzaki}), leading to the conclusion that Hebb's postulate was, for the first time, directly confirmed. However, because of background noise, small changes (1$\%$) would not have been detected. These single synapse tests place an upper limit on the ``per synapse'' error $b$; as expected, the biological error rate seems to be quite low. However, even at a $b$ value of 1$\%$ significant (5$\%$) decreases in performance can occur in the presence of background correlations (Figure 5). As previously noted, the ``error sensitivity'' approaches zero at low error in all cases.

Extremely recent results have examined whether LTP-inducing stimuli at one synapse modify the threshold for inducing LTP at neighboring synapses~\cite{Harvey}. It was argued that while LTP itself does not spread, the threshold-lowering effect does. It seems to us that this distinction is not valid if, as observed, LTP is all-or none. In our model, while the updating process is stochastic and discrete, the weights change quasi continuously and deterministically. Any change in the ``threshold'' in a discrete model would be indistinguishable from the spread of LTP in a continuous model. Such an equivalence is implicitly recognized by Harvey and Svoboda in their argument that their ``threshold-change'' results match the previous Engert-Bonhoeffer results; however, Harvey and Svoboda do not explicitly show the all-or-none character of LTP that presumably obtained in their experiments. We therefore conclude that these results do in fact support our basic premise: that LTP is not completely synapse-specific.

It seems likely that the spread of LTP (or, for discrete updating,
the spread of threshold-lowering) is mediated by the intracellular
diffusion of a mobile factor. In optic tectum, Tao et al. have
obtained evidence that the factor is calcium, since the LTP spread
retracts with maturation in parallel with restriction on calcium
diffusion~\cite{Tao}. Harvey and Svoboda argued that the LTP
threshold-reducing ``crosstalk'' in their experiments cannot be due
to calcium diffusion. This issue, which is related to previous
suggestions that calcium is confined to the spine head by the action
of calcium pumps located in the narrow spine
neck~\cite{Koch}~\cite{Sabatini}, is crucial to the interpretation
of our results, but cannot be discussed in detail here. Our view is
that while calcium localization is very good (perhaps as good at
99$\%$ complete), it could only be perfect if the head were so
isolated from the dendrite that it cannot affect it electrically,
rendering the synapse useless. We also note that since LTP-inducing
protocols rapidly shut down the diffusional coupling of the spine
head to the shaft~\cite{Bloodgood}, presumably by a calcium-mediated
process, it seems unlikely that any factor that calcium releases
could travel further and faster than calcium itself. Furthermore,
any such ``avoidable'' spread of LTP would probably degrade the
utility of Hebbian learning (as shown in the current study).

The only circumstance under which LTP ``leakage'' might actually be beneficial is in the rather unusual scenario envisaged by Mel~\cite{Mel} (and mentioned by Harvey and Svoboda): if the basic computational unit in the brain is not the neuron but a dendritic branch. Thus if each branch can act as a ``neuron-like'' element, one might want a selected subset of inputs to target that branch. This selection could be driven by an initial phase of inspecific Hebbian learning. Subsequently, learning would be restricted to single synapses, so the branch, rather than the neuron, would learn the PC. However, the successful implementation of such a strategy would require a host of undocumented biophysical specializations (such as local backpropagating spikes, calcium and perhaps electrical constrictions at branchpoints etc etc) – in fact all the features that neurons are already known to have. Furthermore, the number of possible inputs to a single branch would be much less than the number of inputs to a neuron. Even more fatally, the putative advantages (there are more branches than neurons) would be impossible to reap, because there could be no corresponding increase in the density of synapses, which our work suggests is the main factor that limits learning, and hence neural information processing.

\subsection{Other covariance-driven models}

It is unlikely that the brain does PCA analysis in the strict sense:
uncorrelated rerepresentation of the correlated activities of a set
of input neurons by projections on the complete set of principal
components, or even a high variance subset. This would not be
efficient (since it ignores the fact that each neuron has similar
channel capacity) nor would it be feasible (since the errors that
would be inevitable if each output neuron is connected to every
input would be intolerable). However, it is likely that the brain
does develop related uncorrelated representations using Hebbian
learning. For example, the center-surround filters seen in retina
and thalamus can be viewed as a local form of
decorrelation~\cite{Atick1}~\cite{Atick2}~\cite{Laughlin}. If an
output neuron learns only from a subset of inputs, this lowers the
error rates (since the synapses can be better separated).  In the
Oja model the normalization is ``multiplicative'', keeping the
weight vector near the hypersphere surface by ``vertical''
adjustments~\cite{Miller}. ``Subtractive'' normalization of the
linear Hebb rule  has also been much
studied~\cite{Miller}~\cite{Dayan}: here all the weights are
adjusted by the same amount, such that the weight vector length is
maintained. Subtractive normalization requires limits on the
weights, and learning points the weight vector into the corner of
the weight space hypercube that is typically closest to the second
eigenvector of ${\bf C}$~\cite{Miller}. We have simulated the effect
of error on this rule (Cox and Adams, unpublished); because the
finite learning rate rule is stochastic, even in the absence of
error the weight vector sometimes points to corners close to the
``correct'' corner, and we found that error increases this tendency
but again in a graceful manner: at all error rates the correct
corner is that which is most occupied. We have only observed sharp
collapses in performance (``error catastrophes'') using nonlinear
Hebbian rules (Cox and Adams, unpublished). The only condition in
which performance drops to chance levels in linear Hebbian models is
the trivial condition when learning becomes completely inspecific
(for example, at very high $n$ values when synapses are very
crowded).

\subsection{Experimental tests}

Is there any evidence that crosstalk of the type we consider here actually does degrade the precision of wiring in the brain? The main problem here is that in no case do we know exactly what the wiring is supposed to be, so it's unclear whether any observed deviation from  ``ideal'' wiring is bug (biology is inaccurate) or feature (biology is cleverer than we are). Here we will focus on a particularly well-studied case, wiring from retina to thalamus, and make some simplifying assumptions. Our main assumptions will be (1) the activity of different retinal ganglion cells is approximately equal and uncorrelated over the ensemble of natural images~\cite{Laughlin}~\cite{Atick1}~\cite{Atick2} (2) although the thalamus is more than a simple relay~\cite{Sherman}, it is approximately so. (3) Hebbian mechanisms control at least the final refinement of thalamocortical wiring. We recognize that none of these assumptions is proven or exact, but they form a useful initial framework for discussion of the role of error. Previous authors have also suggested that some aspects of detailed retinothalamic wiring might result from inevitable wiring  inaccuracy, rather than half-hearted initial ``information processing''. We now justify these assumptions.

\begin{enumerate}

\item Current models suggest that the center-surround organization of ganglion cell RFs reflects a decorrelating strategy in response to limited numbers of limited bandwidth output channels and the statistics of natural images~\cite{Laughlin}~\cite{Atick1}~\cite{Atick2}. This does not mean that ganglion cells are completely uncorrelated, merely that they are as uncorrelated as possible while still transmitting high levels of visual information.

\item In particular, we assume  that if the intercalation of a thalamic relay does alter the spatiotemporal RFs of ganglion cells, it does so only in response to descending influences resulting from initial cortical processing.  Thus the default transformation would be the identity (i.e. spikes would be transmitted one-for-one). Now clearly there is considerable divergence and convergence in retinothalamic wiring~\cite{Sherman}, but our view is that this provides coding flexibility which can only be exploited after the cortex has made a preliminary analysis. Since the retinal representation is in some sense optimized (after all, this is what is sent to the colliculus), any additional ``optimization'' done by thalamus can only be a response to cortical feedback.

\item There is much evidence that many aspects of retinothalamic wiring are achieved by activity-dependent~\cite{Shatz} NMDAR and spike-coincidence based plasticity.

\end{enumerate}

Many lateral geniculate relay cells get only one retinal input, and thus act as simple relays. However although almost all cells get 1 dominant input, many also get 1 or more subsidiary inputs. Hebbian mechanisms can explain the one-input cases if the inputs are all uncorrelated, and it's possible that failure to eliminate subsidiary inputs~\cite{Chen} reflects Hebbian inaccuracy. In particular, the principal and subsidiary inputs RFs tend to be very close together, and could show some low level of correlation; we show above that background correlations and low levels of error are synergistic. Alonso et al.~\cite{Alonso} have argued that such convergence might be useful to provide receptive field diversity, but whether such spatial ``mixing'' is useful would depend on details of spatiotemporal noises and signals.

\section{Conclusions}

Although it is widely appreciated that physics sets ultimate limits
to biology, little attention has been paid to the physical limits to
the process that is of  most interest to humans: learning. The Oja
rule is the simplest and best-studied unsupervised learning rule. It
captures the key point that Hebbian learning is driven by pairwise
correlations (in the form of the input covariance matrix). Not
surprisingly, when the rule is inaccurate, it fails to accurately
learn the expected (and typically most useful) result. Although
the failure is graceful, it can be severe when the patterned
activity driving growth of particular weights is rather weak. We
propose that even though the chemical changes driving Hebbian
learning are largely confined to the synapses where learning is
induced, the very high density of synapses along dendrites means
that significant crosstalk, and therefore somewhat degraded
learning, is inevitable. In future work we hope to show that such
inevitable crosstalk can completely prevent Hebbian learning of
higher-than-pairwise correlations, unless additional interesting
machinery, roughly corresponding to the basic neocortical
microcircuit, is employed.

\clearpage
\centerline{\bf Text S1 -- Stability}
\vspace{5mm}
\indent

 We calculate the Jacobian matrix $Df_{\bf w}$, for a fixed vector ${\bf w}$:

   \begin{lemma} $Df_{\bf w} = {\bf I} + \gamma \left[ {\bf C} - 2{\bf w}({\bf Cw})^{T} - ({\bf w}^{T}{\bf Cw}){\bf I} \right]$
   \end{lemma}

   \proof{ Call $g({\bf w})=({\bf w}^{T}{\bf Cw}){\bf w}$ , so $f({\bf w}) = {\bf w} + \gamma ({\bf Cw} - g({\bf w}))$

$$g_{i}({\bf w}) = ({\bf w}^{T}{\bf Cw})w_{i}$$
If $i \not= j$:
$$\frac{\partial g_{i}}{\partial w_{j}}({\bf w}) =
       \frac {\partial}{\partial w_{j}}(\sum_{k,l}C_{kl}w_{k}w_{l})w_{i}
    =2(\sum_{k} C_{kj}w_{k}) w_{i} \, = \, 2[{\bf Cw}]_{j}w_{i}$$
If $i=j$:
$$\frac{\partial g_{i}}{\partial w_{i}}({\bf w}) =
       \frac {\partial}{\partial w_{i}}(\sum_{k,l}C_{kl}w_{k}w_{l})w_{i}
       +\sum_{k,l}C_{kl}w_{k}w_{l}
       =2(\sum_{k}C_{ki}w_{k})w_{i} +$$
     $$ + {\bf w}^{T}{\bf Cw} = 2[{\bf Cw}]_{i}w_{i} +  {\bf w}^{T}{\bf Cw}$$
So: $$Dg_{\bf w} = 2{\bf w}({\bf Cw})^{T} + ({\bf w}^{T}{\bf Cw}){\bf I}$$}

\vspace{3mm}

  Take now an orthonormal basis ${\cal{B}}$ of eigenvectors of ${\bf C}$ ( with respect to the Euclidean norm $\| \cdot \|$ on $\mathbb{R}^{n}$). Fix a vector ${\bf w} \in {\cal{B}}$. Pick any ${\bf v} \in {\cal{B}} ,{\bf v} \not= {\bf w}$. Call $\lambda_{\bf w}$ and $\lambda_{\bf v}$ their corresponding eigenvalues.

\begin{eqnarray*}
Df_{\bf w}({\bf v})&=&v+\gamma[{\bf Cv} - 2 {\bf w}({\bf Cw})^{T}{\bf v} - ({\bf w}^{T}{\bf Cw}){\bf v}] = \\
&=&{\bf v} + \gamma[{\bf Cv} - 2 {\bf ww}^{T}{\bf Cv} - ({\bf w}^{T}{\bf Cw}){\bf w}]= \\
&=&v + \gamma[\lambda_{\bf v}{\bf v} - 2 {\bf ww}^{T}\lambda_{\bf v}{\bf v} - \lambda_{\bf w}{\bf w}] = (1 - \gamma[\lambda_{\bf w} - \lambda_{\bf v}]){\bf v} \\
\\
Df_{\bf w}({\bf w}) &=& {\bf w} + \gamma [{\bf Cw} - 2 {\bf w}({\bf Cw})^{T}{\bf w} - ({\bf w}^{T}{\bf Cw})] =\\
&=&{\bf w} + \gamma[\lambda_{\bf w}{\bf w} - 2 {\bf ww}^{T}\lambda_{\bf w}{\bf w} - \lambda_{\bf w}{\bf w}]=\\
&=&{\bf w} + \gamma[- 2 \lambda_{\bf w} \| {\bf w} \| {\bf w}] = [1 - 2 \gamma \lambda_{\bf w}]{\bf w}\\
\end{eqnarray*}

\noindent
So ${\cal{B}}$ is also a basis of eigenvectors for $Df_{\bf w}$.

\vspace{10mm}

Our next goal is to generalize this argument for an iteration function that includes errors. The new model introduces an error matrix, ${\bf E} \in {\cal{M}}_{n}(\mathbb{R})$ that has positive entries, is symmetric and equal to the identity matrix ${\bf }I \in {\cal{M}}_{n}(\mathbb{R})$ in case the error is zero. Moreover, we assume that ${\bf EC}$ has strictly positive maximal eigenvalue of multiplicity one.

  $$ f^{\bf E}({\bf w})={\bf w}+\gamma[{\bf ECw}-({\bf w}^{T}{\bf Cw}){\bf w}]$$

  Note that the symmetric, positive definite matrix ${\bf C} \in {\cal{M}}_{n}(\mathbb{R})$ defines a dot product in $\mathbb{R}^{n}$ as:

  $$\langle {\bf v},{\bf w} \rangle_{\bf C}={\bf v}^{T}{\bf Cw}$$
If ${\bf v}$ and ${\bf w}$ are eigenvectors of ${\bf EC}$ corresponding to the eigenvalues $\lambda_{\bf v} \not= \lambda_{\bf w}$, then they are orthogonal with respect to the dot product $\langle , \rangle _{\bf C}$. Indeed:\\

  ${\bf ECv}=\lambda_{\bf v}{\bf v} \: \Rightarrow \: \langle {\bf w},{\bf ECv} \rangle _{\bf C}=\lambda_{\bf v}\langle {\bf w},{\bf v} \rangle _{\bf C}$\\

  ${\bf ECw}=\lambda_{\bf w}{\bf w} \: \Rightarrow \: \langle {\bf v},{\bf ECw} \rangle _{\bf C}=\lambda_{\bf w}\langle {\bf v},{\bf w} \rangle _{\bf C}$\\

\noindent
Hence $\lambda_{\bf v}\langle {\bf v},{\bf w} \rangle _{\bf C}=\lambda_{\bf w}\langle {\bf v},{\bf w} \rangle _{\bf C}$. As $\lambda_{\bf v} \not= \lambda_{\bf w}$, it follows that $\langle {\bf v},{\bf w} \rangle _{\bf C}=0$, hence ${\bf v}$ and ${\bf w}$ are orthogonal with respect to the given dot product.\\

  A fixed point for $f^{\bf E}$ is a vector ${\bf w}=(w_{1}...w_{n})^{T}$ such that ${\bf ECw}=({\bf w}^{T}{\bf Cw}){\bf w}$. In other words, ${\bf w}$ is fixed by $f^{\bf E}$ if and only if it is an eigenvector of ${\bf EC}$ (with corresponding eigenvalue $\lambda_{\bf w}$), normalized such that $\parallel {\bf w} \parallel_{\bf C}=\lambda_{\bf w}$. Clearly, this is possible if and only if $\lambda_{\bf w}>0$.

  $${\bf ECw}=\lambda_{\bf w} {\bf w}, \quad \| {\bf w} \|_{\bf C}=\lambda_{\bf w}$$

\noindent If the multiplicity of $\lambda_{\bf w}$ is one, then ${\bf w}$ is orthogonal in $\langle , \rangle_{\bf C}$ to all other eigenvectors of ${\bf EC}$.\\

  Recall that

  $$Df^{\bf E}_{\bf w}={\bf I}+\gamma[{\bf EC}-2{\bf w}({\bf Cw})^{T}-({\bf w}^{T}{\bf Cw}){\bf I}]$$

  Take ${\bf w}$ to be a fixed point of $f^{\bf E}$. ${\bf w}$ will hence be an eigenvector of ${\bf EC}$, with eigenvalue $\lambda_{\bf w}=({\bf w}^{T}{\bf Cw}){\bf w}>0$. Calculate:

\begin{eqnarray*}
Df^{\bf E}_{\bf w}{\bf w} &=& {\bf w} + \gamma [{\bf ECw} - 2 {\bf w}({\bf Cw})^{T}{\bf w} - ({\bf w}^{T}{\bf Cw}){\bf w}] = \\
&=& {\bf w} + \gamma [- 2 {\bf ww}^{T}{\bf Cw}] =  [1 - 2 \gamma \lambda_{\bf w}]{\bf w} \\
\\
Df^{\bf E}_{\bf w}{\bf v} &=& {\bf v} + \gamma [{\bf ECv} - 2 {\bf ww}^{T}{\bf Cv} - \lambda_{\bf w}{\bf v}] = \\
&=& {\bf v} + \gamma [(\lambda_{\bf v}-\lambda_{\bf w})v -2\langle {\bf w},{\bf v} \rangle _{\bf C}{\bf w}]=(1 - \gamma [\lambda_{\bf w} - \lambda_{\bf v}]) {\bf v}
\end{eqnarray*}

\noindent
for any other eigenvector ${\bf v}$ of ${\bf EC}$ with eigenvalue $\lambda_{\bf v} \not= \lambda_{\bf w}$:\\

   As in the error free case, $Df^{\bf E}_{\bf w}$ has all eigenvalues less than one in absolute value if and only if $\lambda_{\bf w}$
is the principal eigenvalue of ${\bf EC}$ and $\gamma<\frac{1}{\lambda_{\bf w}}$.

\clearpage
\centerline{\bf Text S2 -- Synaptic dynamics}
\vspace{5mm}
\indent

In this Appendix we consider several types of synapse formed on the dendrites of a cell, all made on spines. A ``potential'' synapse refers to any site at which an axon is within a spine length of a dendrite~\cite{Stepanyants}; note that while in then original definition of a potential synapse a fixed spine length was considered, the definition could be generalised to include a distribution of spine lengths, with a suitable redefinition of parameters~\cite{Stepanyants2}; axons and dendrites are assumed to have fixed geometry. We divide the set of potential synapses into the set of existing synapses (which form a fraction $f$ of potential synapses, called the ``filling factor'' by Stepanyants et al.~\cite{Stepanyants}) and the set of ``incipient'' synapses (empty sites where a spine could spontaneously form, to generate an existing synapse. Existing synapses can be either ``silent'' or ``active''. A silent synapse is plastic but has zero strength; it can be promoted to be an active synapse as a result of sufficient past conjoint activity across it, which increments its strength (from zero to one unit in a stochastic manner) depending on the recent history of transynaptic activity (repeated spike pairing or coincidence leading to LTP). Similarly, active synapses can further strengthen in an either continuous or unitary (to 2,3...etc) manner, as a result of further conjoint activity. The history of activity can be expressed either by the accumulation of very small changes (e.g. stochastic insertion of single AMPA receptors as a result of calcium increases) or the accumulation of small calcium signals in a ``register'' (such as CaMKinase)which leads to a larger change (e.g. insertion of a packet of AMPA receptors when a threshold is reached; both mechanisms would show similar long-term behavior.

Active synapses can also discretely decrement (LTD; see Discussion), perhaps eventually resilencing.  Silent synapses can also disappear, recreating an incipient synapse; we assume that the process of forming and removing synapses (which does not change connectional strength) goes on at a steady rate which is similar at all connections (though it may be influenced by the overall level of activity of the pre- and particulalrly post-synaptic neurons). The balance of the unsilencing, silencing, formation and removal processes set the filling fraction, which is typically around 10 $\%$~\cite{Stepanyants}. Stepanyants et al. have pointed out that an $f$ below 1 allows ``anatomical'' plasticity; however, since this comes at the expense of physiological plasticity (strengthening and weakening of existing synapses), there is no net gain; instead we will argue that $f < 1$ lowers the effective error rate, as well as the effective learning rate.

A connection from a given presynaptic neuron to the postsynaptic neuron is made up on average of a number $\alpha$ of synapses, both incipient and existing. Since a feedforward connection in cortex is typically made of about 10 actual synapses, it is comprised of about $10/f$ potential synapses. However, as learning proceeds, the fraction of these potential synapses that comprise an existing connection gradually changes; an input may become anatomically disconnected for extended periods because LTD exceeds LTP, all its synapses become silent, and are then removed; however, such anatomical disconnection does not remove functional connectivity, because continued low-level background generation of silent synapses~\cite{Engert2}~\cite{Trachtenberg}~\cite{Lendvai} can reinstate the anatomical connection for trial periods. The number of existing synapses that comprise a connection is on average $f \times \alpha$, hence the total number of existing synapses along the postsynaptic dendrite is $N=f \times \alpha \times n$.

In most learning models it is assumed that the rate of learning is approximately the same at all connections. A simple way to ensure this would be if each existing synapse has the same ``plasticity'', or responsiveness to transsynaptic activity, and $\alpha$ or $N$ were approximately the same for all inputs. Since weak connections would tend to have fewer existing synapses, they would have slower learning.

\begin{figure}[h!]
\begin{center}
\includegraphics[scale=.3]{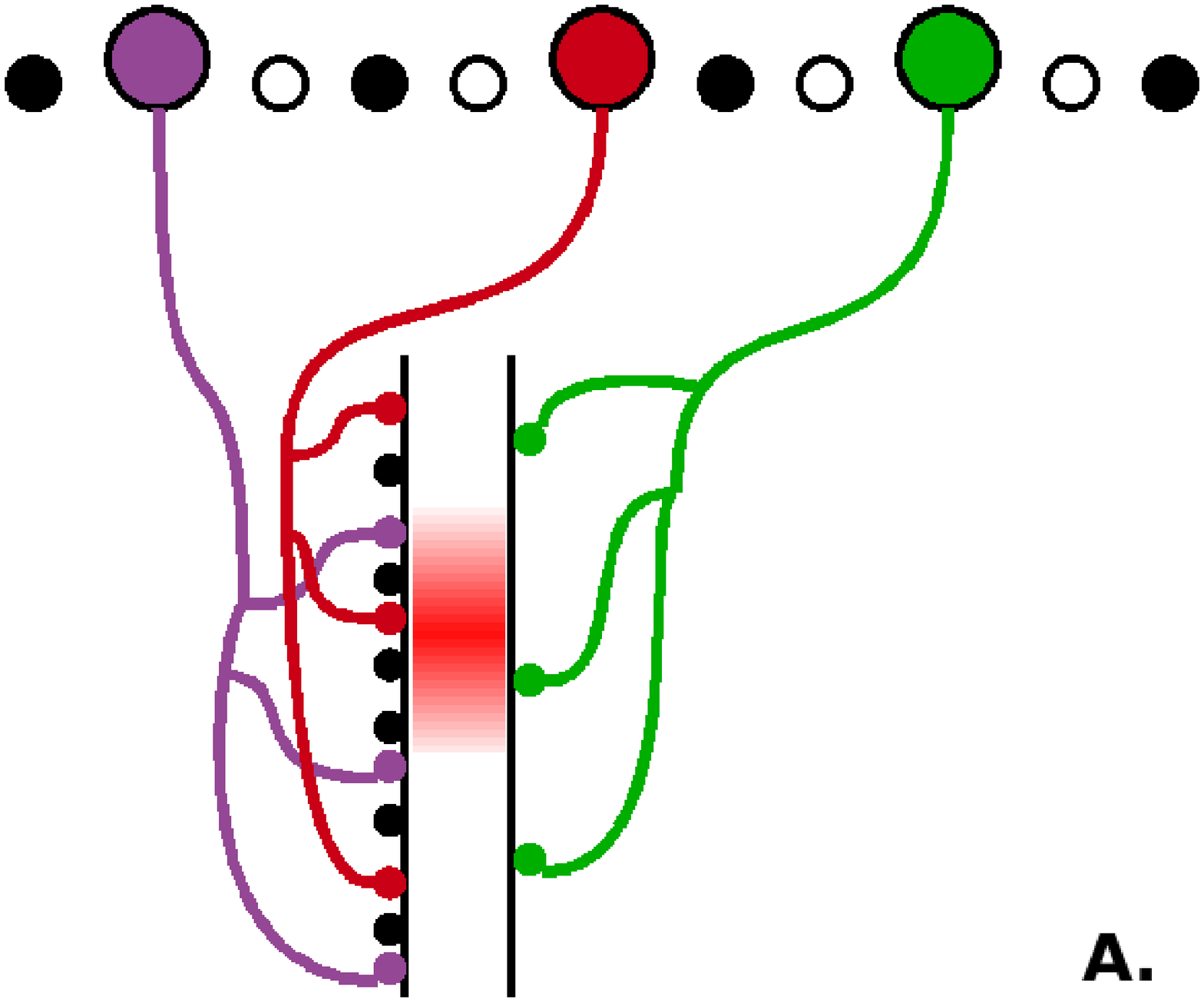} \quad \quad \quad
\includegraphics[scale=.3]{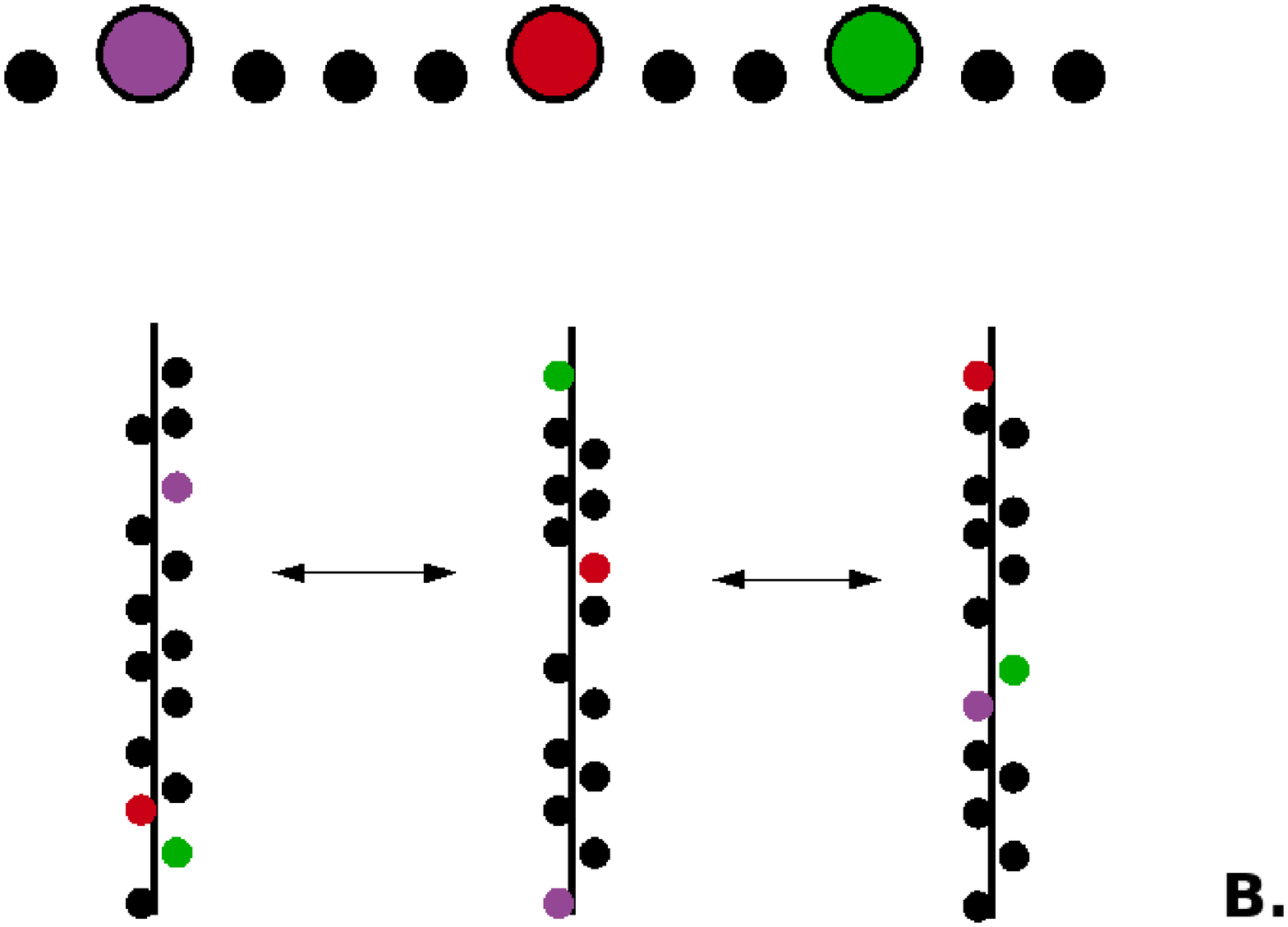}
\end{center}
\caption{\emph{Two routes to 'error-onto-all". Part A (left) shows a dendritic segment that receives potential synapses (on spines, not shown) from a large set of input neurons, some of which are represented schematically as circles on the top row. Connections from three input neurons (colored circles) are shown in more detail; they are each comprised of three existing synapses plus additional incipient synapses (which are not shown). The dendrite beneath the middle red synapse is colored red to indicate the neighborhood over which calcium diffusing from that synapse can act; the action declines exponentially with distances (decreasing red shading). The set of synapses within this neighborhood (and their ordering) is random; in this case a green synapse and a purple synapse are about equally close (distances along the dendritic width are irrelevent), so neither synapse is ``preferred''. However in other cases (eg bottom red synapse) either the green or red connection is favored. Because all possible configuration average out spatially, especially when calcium action spreads further than the average distance between synapses (as shown here), all connections are approximately equally ``close''. The black synapses originate from other neurons (shown in black in the top row). Neurons that do not currently make synapses on this dendrite are shown as dotted circles. Part A corresponds to a mature network, in which many inputs have disconnected as a result of previous learning, and the remaining inputs are quite strong, and have rather stable synapses. Part B (right) shows an earlier situation, when each connections is made of fewer synapses, but more inputs are connected (top row). The bottom row shows three ``snapshots'' of the state of a dendritic segment. In each case the colored inputs make only 1 synapse each, but these shift around on the dendrite (perhaps by first disappearing as a result of normalisation and then reappearing at new locations when a new spine contacts a branch of the same axon that passes nearby). The arrows show that while each connection has a unique configuration of neighbors at any one time, the details get temporally averaged out. A combination of the mechanisms in A and B (mostly B in immature networks and A in mature networks) could ensure that the average error matrix is ``error-onto-all'' throughout protracted learning. }}
\end{figure}

We now consider the neighborhood relations between different connections. We argue that all connections are approximately equivalent, despite the happenstances of particular axodendritic geometries, based on three factors (Figure 1, parts b1 and b2).

\begin{enumerate}

\item If synapses are fixed, but each connection is made up of very many synapses (high $N$) then provided $n$ is not too large all possible neighborhood relations will occur, resulting in spatial averaging (Figure 1b2).

\item If synapses ``turn-over'' as described above, then any given synapse will eventually sample all possible neighbors (Figure 1b1).

\item In both cases, the averaging (spatial or temporal) will be more effective the greater the extent of the neighborhood (shown as red shading in Figure 1b1).

\end{enumerate}

While none of these factors alone would suffice to make all connections equivalent, we suggest that some biologically reasonable combination of them would plausibly do so. Such equivalence would apply both to the electrical effects of connections (where factor 3, reflecting electrical spread, is rather powerful~\cite{Koch}; and to the crosstalk pattern (where factor 3, representing chemical spread, is less powerful). \\

   In this context, consider a cortical volume $V$ containing tightly packed neuronal cells and their processes (see ~\cite{Stepanyants}), a subset of which are organized as a learning network. We calculate the number $N_p$ of potential synapses on a dendrite ${\cal S}$ of length $L$ of the output (postsynaptic) neuron. We approximate ${\cal S}$ by a cylinder with diameter equal to the spine length $2s$, and we divide all axons into little pieces, each oriented at a solid angle $\theta_{i}$ with the axis of ${\cal S}$.

If the distribution of the angle $\theta_{i}$ between $S$ and axonal pieces is isotropic and independent of the particular axon, then the number of potential synapses $N_p$ on ${\cal S}$ is given by a calculation very similar to the one performed in~\cite{Stepanyants}:

  \begin{eqnarray}
  N_p &=& \frac{\pi}{2}\frac{sLl_{a}}{V}n = \alpha n
  \end{eqnarray}

\noindent where $s$ is the spine length, $L$ is the dendritic length, $l_a$ is the average per neuron axonal length, $V$ is the volume that contains the network and $n$ is the network size. Note that $N_p$ is proportional to the size $n$ with a proportionality constant intrinsic to the network.

  In consequence, the number $N$ of existing synapses on ${\cal S}$ will be a fraction $f \times N_p$:

\begin{eqnarray}
  N &=& f \frac{\pi}{2} \frac{sLl_{a}}{V}n= f \alpha n
\end{eqnarray}

  We may reinterprete equation (7) as: the density $\rho$ of existing synapses along a dendritic segment ${\cal S}$ of length $L$ is estimated by:

  $$\rho=f \frac{\pi}{2} \frac{sl_{a}}{V}n$$

\noindent which depends linearly on the size $n$ of the network.

\vspace{5mm}

  Our goal is to estimate the quality factor $Q$ of learning in such a network. Suppose the output cell receives the spike-coincidence signal to strengthen the connection with a prescribed presynaptic axon at a target site $A$. The activation signal will trigger the release of a messenger (such as $Ca^{2+}$) which ideally should appear only at the target site $A$. Instead, the messenger is not completely locally contained and diffuses down the dendritic spine. A small portion $a_1$ of it ($a_1 \sim 10^{-2}$) will reach the dendritic ``cable'' and leak along it in both directions, such that its concentration at a distance $x$ from the target spine follows a function $g(x)=e^{-x/\lambda}$. This messenger could spread along the dendrite to some arbitrary site $B$, which represents an existing (either active or silent) synaptic site. Some proportion of it $a_2 \sim 10^{-1}$ will travel up the spine, arriving at the spine head, where is could update  the synaptic mechanism (i.e., either activate a silent synapse or strengthen an already active synapse).

   We compute the probability for a messenger to get from the target site $A$ to site $B$, situated at distance $x$ from $A$ along the dendrite as: $ a \, e^{-x/\lambda}$, where $a=a_1a_2 \sim 10^{-3}$. The probability for the update to be activated at site $B$ depends on a power $h$ of the messenger concentration:

  $$ p_{x}= a^{h} e^{-hx/\lambda}$$

  If we assume that the distribution of existing synapses along the dendrite is homogeneous, then the probability for and arbitrary synapse to be activated anywhere along the length $L$ dendrite is:

  $$b= \frac{1}{L} \int_{0}^{L} p_{x} dx= \frac{1}{L} \int_{0}^{L} a^{h} e^{\frac{-hx}{\lambda}} dx = \frac{a^{h} \lambda}{hL} (1-e^{\frac{-hL}{\lambda}})$$

   Suppose now that we knew the positions $x_{1},...x_{N}$ of the $N$ existing synapses along the postsynaptic dendrite. The probability that messenger diffusion from site $A$ affects exactly the synapses at positions $x_{j_{1}},...,x_{j_{k}}$ is

   $$ \sum_{1 \leq j_{1}<...<j_{k} \leq N} {p(x_{j_{1}})...p(x_{j_{k}}) [1-p(x_{j_{k+1}})] ... [1-p(x_{j_{N}})]}dx_{j_{1}} dx_{j_{N}}$$

   Hence the probability (over all possible placements of existing synapses along the dendrite) to update exactly $k$ of these synapses is:

   \begin{eqnarray*}
   && \frac{1}{L^{N}} \int_{[0,L]^{N}} \!\!\! \sum_{1 \leq j_{1}<...<j_{k} \leq N} {p(x_{j_{1}})...p(x_{j_{k}}) [1-p(x_{j_{k+1}})] ... [1-p(x_{j_{N}})] dx_{1}...dx_{N}}=\\
    &=& \frac{1}{L^{N}} \sum_{1 \leq j_{1}<...<j_{k} \leq N} \int_{[0,L]^{N}} {p(x_{j_{1}})...p(x_{j_{k}}) [1-p(x_{j_{k+1}})] ... [1-p(x_{j_{N}})] dx_{1}...dx_{N}}=\\
    &=& \frac{1}{L^{N}} \sum_{1 \leq j_{1}<...<j_{k} \leq N} \int_{[0,L]} p(x_{j_{1}})dx_{j_{1}} ... \int_{[0,L]} p(x_{j_{k}})dx_{j_{k}} \\
    && \quad \quad \int_{[0,L]} [1-p(x_{j_{k+1}})]dx_{j_{k+1}} ... \int_{[0,L]} [1-p(x_{j_{N}})] dx_{j_{N}}=\\
    &=& {{N}\choose{k}} \left( \int_{0}^{L} p(x)dx \right)^{k} \left( 1-\int_{0}^{L} p(x)dx \right)^{N-k}={{N}\choose{k}} b^{k}(1-b)^{N-k}
   \end{eqnarray*}

\noindent where ${{N}\choose{k}}=\frac{N!}{k!(N-k)!}$ is the number of all possible combinations to choose $k$ elements out of a set of $N$.

    If $k$ sites are additionally activated, it means that $k+1$ synapses have been overall updated (including the target site at $A$). In order to obtain the desired degree of strengthening required by the learning rule, normalization decides at random which one of these $k+1$ changes should be applied and discards the other ones.

   With this normalization, the probability of the synapse that wins the update to be the correct target synapse is equal to:

   $$ \frac{1}{k+1} \; \frac{N!}{k!(N-k)!} b^{k}(1-b)^{N-k} $$

   We calculate the probability that the system turns and keeps on the desired target synapse, no matter how many other wrong synapses have fickered on and off in the normalization process. In this {\bf ``discrete'' model}, the ``quality'' $Q$ of the information transfer will be:

   \begin{align}
   Q =& \sum_{k=0}^{N} \frac{1}{k+1} \; \frac{N!}{k!(N-k)!} b^{k}(1-b)^{N-k} \nonumber \\
     =& \frac{1}{b(N+1)} \; \sum_{k=0}^{N} \frac{(N+1)!}{(k+1)!(N-k)!} b^{k+1}(1-b)^{N-k} \nonumber \\
     =& \frac{1}{b(N+1)} \; \sum_{j=1}^{N+1} \frac{(N+1)!}{j!(N+1-j)!} b^{j}(1-b)^{N+1-j} \nonumber \\
     =& \frac{1-(1-b)^{N+1}}{b(N+1)}
   \end{align}

For further calculations, $Q$  can be approximated for small values of $b$ by $Q=(1-b)^{N/2}$, which has the same linear expansion around $b=0$ as the original expression. This is one approximation of the discrete model that we use to simplify analysis. It is applicable in our context, since learning operates with small error values.

Another useful approximation is given by an average treatment of the same problem, as follows. The expected number (over a large number of epochs) of existing sites updated along the dendrite ${\cal S}$ is the probability $b$ of a site to turn on anywhere multiplied by the number of such sites. The total error {\Large{$\epsilon$}} is the ratio of the number of sites expected to erroneously update and the expected number of all affected sites: {\Large{$\epsilon$}}={\Large{$\epsilon$}}$(N,b)=\displaystyle{\frac{Nb}{Nb+1}}$, hence $\displaystyle{Q=\frac{1}{Nb+1}}$. This average estimate, which we call the {\bf continuous model}, offers a good approximation for (8) for both small and large values of $b$. As

$$ \lim_{N \to \infty} \frac{[1-(1-b)^{N+1}](1+Nb)}{b(N+1)} = 1 \, , \text{ for all }\, b \in [0,1] $$

\noindent the two expressions agree in the limit when $N \to \infty$, since in this case the number of synapses becomes unlimited.

Figure 13 shows the relationship between the exact expression for $Q$ given by the discrete model and the two approximations we use.

\begin{figure}[h!]
\begin{center}
\includegraphics[scale=0.75]{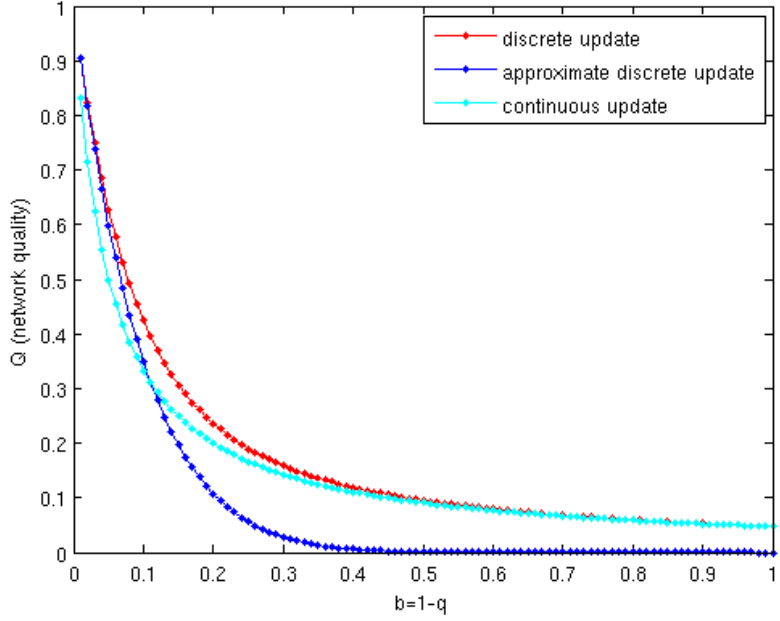}
\caption{{Comparison between the two synaptic error models. For  $N=20$ existing synapses, we show the dependence of the overall network learning quality with respect to the error factor $b$ in three cases: continuous model $Q=\frac{1}{Nb+1}$, approximate discrete model $Q=(1-b)^{N/2}$ and exact discrete model $Q=\frac{1-(1-b)^{N+1}}{(N+1)b}$.}}
\end{center}
\end{figure}

Since $N=\alpha \times f \times n$, we may choose for simplicity the units in the continuous model to be such that $\alpha \times f=1$; $Q$ will then depend on the network size $n$ as $\displaystyle{Q=\frac{1}{nb+1}}$. In the approximate discrete model, we make $\alpha \times f=2$, so that the dependence on $n$ becomes: $Q=(1-b)^n=q^n$. \\

We now examine how maximum error sensitivity depends on $q=1-b$ and $n$, in the discrete and continuous cases. \\

\noindent \underline{\bf Continuous model:}

$$Q=\frac{1}{n(1-q)+1}$$

We calculate the second derivative of $h$. Notice that this
discussion applies for all values of $q=1-b \in [0,1]$ except
$q=q_0=1/n$, where $Q=\epsilon$ and ${\bf EC}$ becomes singular.\\

$\displaystyle{h''(1)=\frac{2n^3\lambda(n-2)}{(\lambda-1)^2(n-1)\sqrt{n-1}}
> 0}$\\

$\displaystyle{h''\left( \frac{1}{n} \right)=\frac{-2n^2(\lambda-1)}{(n+\lambda-1)
\sqrt{n-1}} \left[ \frac{n\lambda}{(n-1)(n+\lambda-1)} +1 \right] < 0}$\\

\noindent There is an inflection point for $h$ in the interval
$\displaystyle{\left(\frac{1}{n},1 \right)}$

\noindent \underline{\bf Discrete model:}

$$Q=q^n$$

The following applies only for $q \neq q_0$, where again $q_0$ is
the value that makes ${\bf E}$ singular, in this case
$q_0=1/\sqrt[n]{n}$.\\

$\displaystyle{h''(1)=\frac{2n^3\lambda(n-2)}{(\lambda-1)^2(n-1)\sqrt{n-1}}
> 0}$\\

${\displaystyle h''\left( \frac{1}{\sqrt[n]{n}}
\right)=-\frac{\lambda-1}{\sqrt{n-1}} \;
\frac{\mu''(\mu-1)-2(\mu')^{2}}{(\mu'-1)^{2}} <0}$\\

\noindent There is an inflection point for $h$ trapped in the interval
$\displaystyle{\left( \frac{1}{\sqrt[n]{n}},1 \right)}$.\\

\clearpage
\centerline{\bf Text S3 -- The output performance for correlated inputs}

\vspace{5mm}

\centerline{\bf Model 1 -- high covariance on one pair}

\vspace{5mm}

In matrix form, ${\bf E}=\epsilon {\bf M}+(1-n \epsilon){\bf I}$ and ${\bf C}=\xi {\bf M}+(\lambda-\xi){\bf A}+(1-\xi){\bf I}$, where the matrices: ${\bf M}=(m_{ij})$, with $m_{ij}=1, \forall i,j \in \overline{1,n}$ and ${\bf A}=(a_{ij})$, with $a_{12}=a_{21}=1$ and $a_{ij}=0$, otherwise. Therefore:

$${\bf EC}=[\epsilon+\xi+(n-2)\epsilon
\xi]{\bf M}+(1-n \epsilon)(\lambda-\xi){\bf A}+\epsilon(\lambda-\xi){\bf MA}+(1-n \epsilon)(1-\xi){\bf I}$$

\noindent and the characteristic polynomial of ${\bf EC}$ is:

$$\lvert {\bf EC}-x{\bf I} \rvert=[(1-n \epsilon)(1-\lambda)-x][(1-n \epsilon)(1-\xi)-x]^{n-3} p(x)$$

\noindent where:
\begin{eqnarray*}
p(x)&=&[(1-n \epsilon)(1-\lambda)-x][(1-n \epsilon)(1-\xi)-x+(n-2)(\epsilon+\xi-\epsilon \xi)]\\
&+&2((1-n \epsilon)(1-\xi)-x)[\epsilon+\xi-\epsilon \xi +
\epsilon(\lambda-\xi)]
\end{eqnarray*}

\noindent So the eigenvalues of ${\bf EC}$ are: $x=(1-n \epsilon)(1-\xi)$
with multiplicity one, $x=(1-n \epsilon)(1-\xi)$ with
multiplicity $(n-3)$ and two more obtained from the solutions of the
quadratic equation:

\begin{equation}
z^2+[(1-(n-2)\epsilon)(\lambda-\xi)+n(\epsilon+\xi-\epsilon
\xi)]z+(1-n
\epsilon)(\lambda-\xi)(n-2)(\epsilon+\xi-\epsilon \xi)=0
\end{equation}

\noindent where we substituted $z=(1-n \epsilon)(1-\xi)-x$. Notice that,
for $\epsilon=0$, the equation becomes:

\begin{equation}
z^2+[\lambda+(n-1)\xi]z+(n-2)\xi(\lambda-\xi)=0
\end{equation}

The maximal eigenvalue of ${\bf EC}$ can be computed as
$\mu_{\bf EC}=(1-n \epsilon)(1-\xi)-z^{-}_{\bf EC}$ and the maximal
eigenvalue of ${\bf C}$ is $\mu_{\bf C}=(1-\xi)-z^{-}_{\bf C}$, where $z^{-}_{\bf EC} < 0$ is the smaller root of (5) and $z^{-}_{\bf C} < 0$ is the
smaller root of (6). Moreover, upper and lower bounds for $z^-_{EC}$ give us corresponding estimates for $\mu_{\bf EC}$:\\

\noindent $\mu_{\bf EC} \geq (1-n \epsilon)(1-\xi)+(1-(n-2)\epsilon)(\lambda-\xi)$\\

\noindent $\mu_{\bf EC} \geq (1-n \epsilon)(1-\xi)+n(\epsilon+\xi-\epsilon
\xi)=1+(n-1)\xi$ and \\

\noindent $\mu_{\bf EC} \leq
(1-n \epsilon)(1-\xi)+(1-(n-2)\epsilon)(\lambda-\xi)+n(\epsilon+\xi-\epsilon \xi)$\\

\noindent Notice that $\mu_{\bf EC}$ goes to $\infty$ as $n \to \infty$,
but the bounds with respect to the other parameters are quite strict
for $\lambda-\xi$ small.

\vspace{7mm}

\centerline{\bf Model 2 -- uniform pairwise covariance}

\vspace{5mm}

 In compact matrix notation, ${\bf C}=(\lambda-1){\bf A}+(1-\xi){\bf I}+\xi {\bf M}$. The characteristic polynomial of ${\bf EC}$ will be:

$$\lvert {\bf EC}-x{\bf I}
\rvert=[(1-n\epsilon)(1-\xi)-x]^{n-2}p(x)$$

\noindent where

\begin{eqnarray*}
p(x)&=&[(1-n \epsilon)(\lambda-\xi)-x][((1-n \epsilon)(1-\xi)-x)+(n-1)(\xi + \epsilon(1-\xi))]\\
&+&[(1-n \epsilon)(1-\xi)-x][\xi +\epsilon(\lambda-\xi)]
\end{eqnarray*}

Hence ${\bf EC}$ has an eigenvalue $x=(1-n\epsilon)(1-\xi)$ with
multiplicity $n-2$ and two others, given by the remaining quadratic
equation (If we assume again that $Q>\epsilon$, the remaining equation has two distinct real roots) . For simplifying notation, call again
$z=(1-n\epsilon)(1-\xi)-x$ and consider the two roots $z^+_{\bf EC}>z^-_{\bf EC}$ of the new equation :

\begin{equation}
z^2+z[(\lambda-1)(1-(n-1)\epsilon)+n(\xi+\epsilon(1-\xi))]+(n-1)(1-n \epsilon)(\lambda-1)(\xi+\epsilon(1-\xi))=0
\end{equation}

The unique largest eigenvalue of $EC$ is $\mu_{\bf EC}=(1-n\epsilon)(\lambda-\xi)-z^-_{\bf EC}$. As upper and lower bounds for the $\mu_{EC}$ we get:\\

\noindent $\mu_{\bf EC} \geq (1-n \epsilon)(\lambda-\xi)$\\

\noindent $\mu_{\bf EC} \geq 1+ (n-1)\xi +\epsilon(\lambda-1) $\\

\noindent $\mu_{\bf EC} \leq (1-n \epsilon)(\lambda-\xi) + 1+ (n-1)\xi +\epsilon(\lambda-1)$

\end{document}